\documentclass[prb,amsmath,amssymb,twocolumn,showpacs]{revtex4-1}
\usepackage{graphicx}
\usepackage{tabularx}
\usepackage{color}
\usepackage{amsmath}
\usepackage{mathtools}
\newcommand{\be}{\begin{equation}}
\newcommand{\ee}{\end{equation}}
\newcommand{\bea}{\begin{eqnarray}}
\newcommand{\eea}{\end{eqnarray}}
\usepackage{dcolumn}
\usepackage{hyperref}
\usepackage{bm}
\usepackage{epsf}
\usepackage{subfigure}
\usepackage{epstopdf}%
\usepackage{amsfonts}
\usepackage{braket}
\usepackage{cancel}

\begin{document}

\title{Coherence and decoherence in quantum absorption refrigerators}
\author{Michael Kilgour and Dvira Segal}

\affiliation{
Department of Chemistry and Centre for Quantum Information and Quantum Control,
University of Toronto, 80 Saint George St., Toronto, Ontario, Canada M5S 3H6
}
\date{\today}

\begin{abstract}
Absorption refrigerators transfer thermal energy from a cold reservoir to a hot reservoir 
using input energy from a third, so-called work reservoir. 
We examine the operation of quantum absorption refrigerators 
when coherences between eigenstates survive in the steady state limit.
In our model, the working medium comprises a discrete, four-level system.
We manifest that eigenbasis quantum coherences within this system 
generally suppress the cooling current in the refrigerator, while minimally affecting   
the coefficient of performance (cooling efficiency).
We rationalize the behavior of the four-level refrigerator by studying two, three-level model systems
for energy transport and refrigeration. 
Our calculations further illuminate the shortcomings of secular quantum master equations, 
and the necessity of employing dynamical equations of motion that retain couplings between population and coherences.
\end{abstract}

\maketitle

\section{Introduction}
\label{sec-intro}




%
Nanoscale classical and quantum heat machines (QHMs)
can be constructed out of few ions \cite{QARE},
atomic-like systems such as natural defects in diamond \cite{Poem},
and even one  atom \cite{singleatom}, to convert input energy into mechanical, 
optical, electrical, or thermal work.
By utilizing non-classical resources, quantum heat machines
may even perform beyond the Carnot limit \cite{squeezeE}.
Paired with exciting real-world applications, QHMs present a rich playground to 
investigate fundamental questions at the intersection of 
thermodynamics and quantum dynamics \cite{kos13,Goold,review1,kos18,Poem,Kurizki_2018,Kurizki_2018_2}. 




Of obvious importance in the study of QHMs 
are questions over the thermodynamic consistency of the employed open quantum systems methodologies.
Typically, projection operator methods are employed in this area \cite{Breuer,Nitzan}, often relying 
on a perturbative expansion in the system-bath coupling energy, a Markovian (fast-bath) approximation, and the
neglect of coherences in the working system through the application of the so-called secular (or ``rotating-wave'') approximation.
Indeed, it has been shown that certain popular quantum master equation approaches, such as the local-basis Lindblad method,
may not respect the laws of thermodynamics, e.g. allowing the heat current to flow against the temperature gradient
\cite{Michel_2007,Wuheat,Kosloff_2014, Brunner_2017,Adesso_2017}.
This issue is vexing enough without considering that it is not generally easy nor necessarily possible to 
prove that a certain approximate approach will respect thermodynamic laws.


Studies of coherence and decoherence in condensed phases are ubiquitous in the open quantum systems literature.  
Relevant to our work are the extensive discussions around the role of quantum 
coherences in energy transfer \cite{Scholes,Miller,Scholes17} 
and in the operation of QHMs 
\cite{Scully_2010,Scully_2011,Scully_2012,Chin_2013,Uzdin,Kosloff_2016,Cao_2018,seah, zhang,Kurizki_2015}.
For a recent review from a resource theory perspective, see Ref. \cite{Plenio_2017}.
Questions around the role of coherences in QHMs and the thermodynamic consistency of 
methods are in fact intimately linked. 
For example, global-secular quantum master equation approaches,
such as the Gorini-Kossakowski-Lindblad-Sudarshan (GKLS) equation \cite{Pascazio_2017,Breuer}, 
neglect the effect of coherences on device operation. 
Local-secular quantum master equations can re-include some of the effects of coherence,
via a change of basis, coming at the cost of strict thermodynamic consistency  
\cite{Kosloff_2014,Adesso_2017,Brunner_2017}. 
On the other hand, {\it nonsecular} quantum master equations of the Redfield type \cite{Nitzan} 
cannot in general be guaranteed to provide fully positive populations
due to issues concerting initial system-bath correlations
\cite{Haake_1996,Oppenheim_1992,Pechukas_1994}, and in general are cumbersome to treat analytically.
Furthermore, the Redfield equation may be exercised in the local-site basis,
resulting in an improper long-time (equilibrium) state for the system \cite{Silbey,SegalET00,Kulkarni16}. 


In this work, we address the interplay of quantum coherences and decoherence processes
in an analytically tractable model for 
a quantum absorption refrigerator (QAR) \cite{reviewARPC14}. 
We  emphasize that we are considering device operation only in the weak system-bath coupling limit. 
Performance of strongly-coupled (system-bath) devices is a pursuit of a significant importance, 
but due to methodological constraints it is outside the scope of the present paper.

A QAR is a continuous-cycle, nondriven heat machine. It extracts energy from a cold reservoir 
and dumps it to a hot reservoir, assisted by energy absorbed from a so-called work reservoir.
The simplest version of such a device, the three-level quantum absorption refrigerator (3lQAR) \cite{DuBois_1959}, 
has been explored in detail in the weak system-bath coupling limit.
Its operating characteristics, such as power and efficiency, are easy to derive when 
employing Markovian quantum master equations, see e.g. Refs. \cite{Levy12,joseSR}.
Internal system coherences disappear in the 3lQAR in the steady state limit, 
and the cooling performance can approach the Carnot bound.
The basic model and its extensions were recently discussed e.g. in Refs. \cite{PopescuPRL,Linden11,Popescu12}.
Recent studies on QARs looked at physical realizations \cite{plenio}, 
the impact of internal leaks and dissipation \cite{jose15}, 
and the roles of topology \cite{AlonsoNJP}, internal couplings \cite{seah}, and strong system-bath coupling effects \cite{Anqi}.
Other questions of interest include the 
cooling performance of the QAR at maximum power \cite{jose13,Alonso14,joseSR},
and the behavior of its current fluctuations \cite{QARfluct}.

Here, in order to probe the role of steady state coherences in the operation of QARs,
we build a four-level quantum absorption refrigerator (4lQAR);
in our design, the four-level working fluid is coupled to three thermal reservoirs 
(work, hot and cold), as well as to
to an additional, purely decohering bath. 
This minimal model allows us to interrogate the roles of internal coupling strength, 
coherence and decoherence in the device operation.
Furthermore, using the 4lQAR model we identify and diagnose the shortcomings of 
secular master equations in the context of QHMs.

To help us understand the operation of the 4lQAR, we analyze two simpler, three-level model systems: 
the V energy transfer system (VETS) and the 3lQAR. 
The VETS, a generalization of the well-studied V-system \cite{Scully06,Brumer14,Brumer15,Brumer16}, 
describes energy exchange between two thermal reservoirs. 
It will give us insight on methodological issues and the role of coherence in the transport behavior.
The 3lQAR forms the basic template for a QAR, which we use to examine the behavior of the more compound 4lQAR.



Previous studies have looked at coherence effects in QHMs, see for example \cite{Scully_2010,Scully_2011,Scully_2012,Chin_2013,Uzdin,Kosloff_2016,Cao_2018,seah,zhang,Kurizki_2015}.
Here, we provide a rigorous and thorough analysis of the problem within a model system,
by studying the interplay between a wide range of internal system couplings and decoherence effects, 
which largely determine the device performance. 
Our examination is grounded in a microscopic quantum mechanical equations of motion derived without heuristics.
Further, discussions of methodological aptitude of secular master equations have largely focused on problems of energy transport  and especially recently, on the local vs. global issue rather than the 
secular vs. nonsecular problem
\cite{Kulkarni16,Adesso_2017,Brunner_2017}. 
In our discussion below we address these questions in the context of both energy transfer and heat to work conversion.

Briefly, the central observations of this work are:
(i) The presence of steady state coherences in the 4lQAR model degrade the cooling current, but not the efficiency, which in 
fact is largely unaffected by internal coherences.
(ii) The cooling performance of the model can be explained based on the behavior of its building blocks (the VETS and 3lQAR),
allowing us to identify key mechanisms in compound models.
(iii) The popular secular approximation provides nonphysical predictions for energy transfer and cooling when applied
outside its regime of applicability. While this observation is not overly surprising, it emphasizes
the importance of a judicious application of open quantum system techniques in the area of quantum thermodynamics.

The paper is organized as follows. We describe the three models, VETS, 3lQAR and 4lQAR in Section \ref{sec-model}.
The performance of these models and possible realizations are presented in  Section \ref{sec-res};
details of the derivations are left to Appendices A-C.
We summarize our work in Sec. \ref{sec-summ}.

\section{Models}
\label{sec-model}

The total Hamiltonian of an open quantum system coupled to multiple (counted by $\alpha$) reservoirs
includes a system Hamiltonian  $\hat H_s$,
thermal reservoirs $\hat H_{\alpha}$, and system-bath coupling terms 
\bea
\hat H= \hat H_s + \sum_{\alpha} \hat{H}_{\alpha} + \sum_{\alpha} \hat S_{\alpha} \otimes \hat B_{\alpha}.
\eea
For simplicity, the reservoirs are assumed to comprise collections of independent 
harmonic oscillators with creation operator $\hat b^\dagger_{q,\alpha}$ of mode  $q$
with frequency $\omega_{q,\alpha}$ in the $\alpha$ bath,
\bea
\hat{H}_{\alpha}=\sum_q\hbar\omega_{q,\alpha}\hat{b}^\dagger_{q,\alpha}\hat{b}_{q,\alpha}.
\eea
The bath operator $\hat B_{\alpha}$ describes displacements of bath oscillators from equilibrium,
\bea
\hat{B}_\alpha=\sum_q\lambda_{q,\alpha}\left(\hat{b}^\dagger_{q,\alpha}+\hat{b}_{q,\alpha}\right).
\eea
$\hat S_{\alpha}$ is a system operator, which is system specific.
Before studying the 4lQAR, we examine two related, simpler, three-level models.
The V energy transfer system describes energy exchange between two heat baths
mediated by a three-level quantum system.
The three-level QAR operates with null steady state coherences. 
Using a weak-coupling quantum master equation,
we study energy transport and refrigeration in these two models. Equipped with this background, we
are able then to rationalize the properties of the four-level QAR.


\begin{figure}[htpb]
\includegraphics[width=8cm]{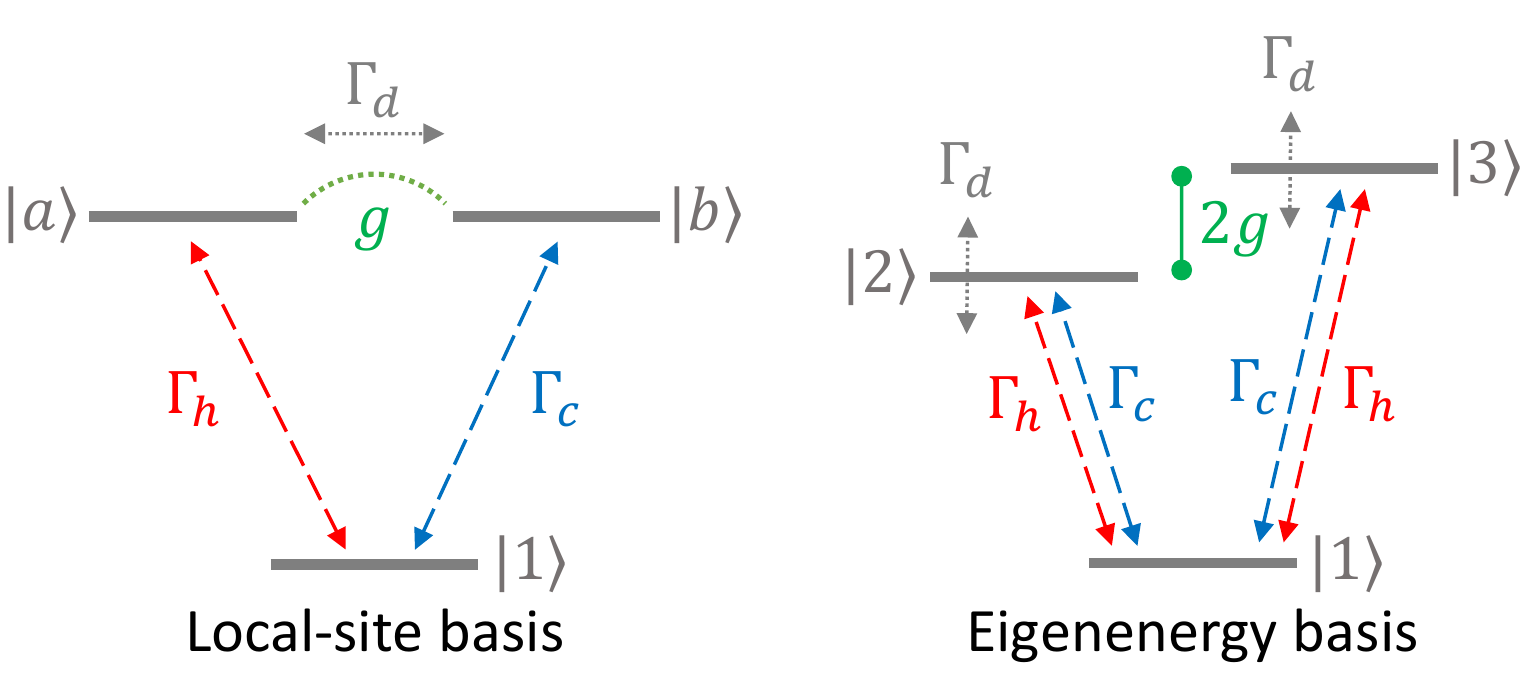}
\caption{
Diagram of the NEVS in the local site basis (left) and in the energy basis (right). 
The red and blue arrows represent energy exchange processes with the hot and cold baths, respectively.
The gray arrow depicts the effect of the decoherence bath.
}
\label{fig:VETS}
\end{figure}

\subsection{V energy transfer system}

The V and $\Lambda$-type systems,  with three atomic or molecular levels 
interacting with coherent or incoherent reservoirs
(representing e.g. electromagnetic radiation fields)
have been intensively investigated in quantum optics. 
In particular, the prospect of generating long-lived
coherences in a quantum system has attracted much attention, with applications to 
lasing, photovoltaic devices and light harvesting systems \cite{Scully06,Brumer14,Brumer15,Brumer16}.
While the dynamics of populations and coherences in V and $\Lambda$ systems has been investigated in detail, so far,
few studies considered the nature of energy transport across such systems \cite{Wu15}. 

We study the behavior of the V system out of equilibrium in the steady state limit.
The model includes three levels and three reservoirs, $\alpha=h,c,d$, see Fig. \ref{fig:VETS}.
The hot ($h$) and cold ($c$) thermal baths exchange energy---mediated by the system.
The so-called dephasing ($d$) bath is responsible for pure decoherence effects.
In the site, ``local" basis, the quantum system includes a ground state $|1\rangle$ 
and two excited states, $|a\rangle$ and $|b\rangle$, which are made degenerate at energy $\theta$.
These states are coherently coupled, with a coupling strength $g$, 
\bea
\hat H_s&=&\epsilon_1 |1\rangle \langle 1 | +\theta  \left( |a\rangle \langle a | +|b\rangle \langle b | \right)
\nonumber\\
&+& g \left(|a\rangle \langle b | +|b\rangle \langle a | \right).
\label{eq:HsNEVS}
\eea 
The coupling energy $g$ can be made arbitrarily large.
In what follows, we set the reference energy at $\epsilon_1=0$. 
The system $\hat S$ operator mediates energy exchange processes and decoherence effects from the environment,
\bea
\hat S_{h}&=& |1\rangle \langle a| +h.c.,  \,\,\,\,   
\hat S_{c}= |1\rangle \langle b| +h.c.
\nonumber\\
\hat S_{d}&= &|a\rangle \langle b| +h.c.
\eea
In the absence of the $d$ bath, energy is transferred between the hot and cold baths as long as $g\neq 0$. 
In the site basis, we picture energy flow as follows: The hot bath excites the $|a\rangle$ site, which
transfers population to  the $|b\rangle$ site through the coherent coupling $g$. 
Since the population of $|b\rangle$---relative to the ground state---exceeds the equilibrium 
value as dictated by the cold bath to which it is coupled, energy is released to the cold bath.
Since the excited states are degenerate, the $d$ bath does not provide energy to the system (in the present weak coupling approach). 
In the system energy basis, the Hamiltonian and the $\hat S_{\alpha}$ operators transform into 
\bea
\hat H_s=   \epsilon_1 |1\rangle \langle 1|+  \left( \theta-g  \right)  |2\rangle \langle 2 | +
\left(\theta+g \right) |3\rangle \langle 3|,
\label{eq:HNEVSd}
\eea
and
\bea
\hat S_c  
&=&\frac{1}{\sqrt{2}} \left(    |1\rangle \langle 3| - |1\rangle \langle 2|  +h.c. \right)
\nonumber\\
\hat S_h&=&  \frac{1}{\sqrt{2}} \left(    |1\rangle \langle 2| + |1\rangle \langle 3|  +h.c. \right)
\nonumber\\
\hat S_d&=& |3\rangle \langle 3| -  |2\rangle \langle 2|.
\label{eq:SNEVS}
\eea 
In this picture, both excited states are coupled to the ground state via both the hot and cold baths.
One may view the model now as having two spins sharing a common ground state, 
with each spin individually but not independently facilitating energy transfer between the two heat baths.
It is also clear now that the $d$ bath is responsible for decoherence, without energy exchange.
As we shall see, since both effective spins
are coupled to the same heat baths, coherences are generated in the system eigenbasis.

\subsection{Three-level QAR}

\begin{figure}[htpb]
\includegraphics[width=4cm]{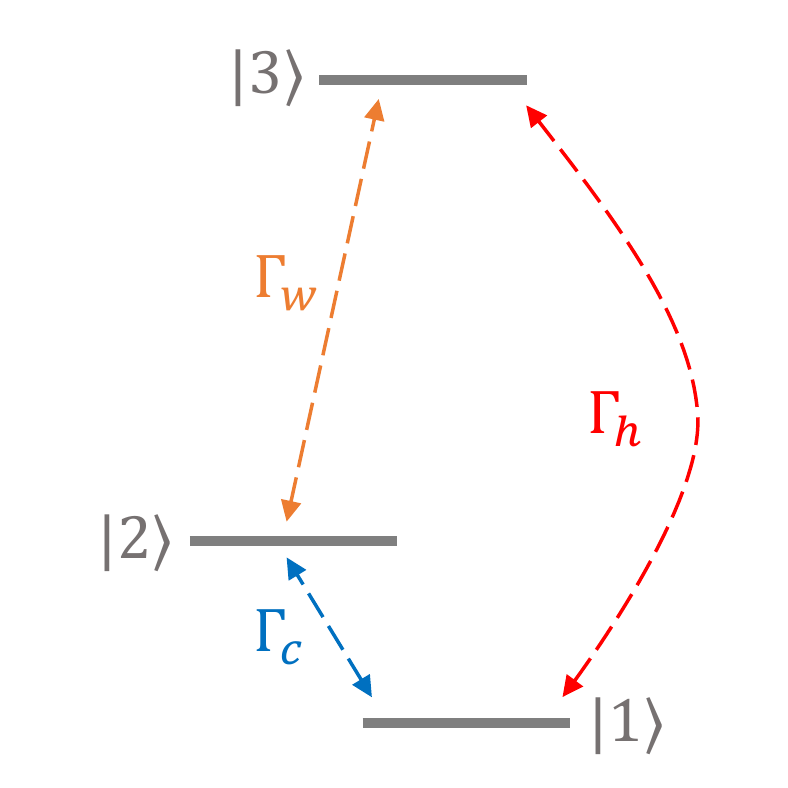}
\caption{
Diagram of a three-level QAR. 
The red, blue and orange arrows represent energy exchange 
processes with the hot, cold and work heat baths.
}
\label{fig:3lQAR}
\end{figure}

A simple design of an autonomous quantum absorption refrigerator
consists of a three-level system as the working medium and
three independent thermal reservoirs \cite{reviewARPC14,joseSR}, see Fig. \ref{fig:3lQAR}.
Each transition between a pair of levels is coupled to one of the three
heat baths, $c$, $h$ and $w$, where $T_w>T_h>T_c$. 
In the steady state limit, the work bath provides energy to the system,
allowing the extraction of energy from the cold bath, to be dumped into the hot reservoir. The opposite
heating process from the hot bath to the cold can be controlled 
by manipulating the frequencies of the system.
The three-level QAR is described by the system Hamiltonian  ($\epsilon_1=0$)
\bea
\hat H_s=\theta_c  |2\rangle \langle 2|+ \theta_h  |3\rangle \langle 3|,
\eea
and we set $\theta_w=\theta_h-\theta_c$.
The $\hat S_{\alpha}$ operators are
\bea
\hat S_{c}&=&|1\rangle\langle 2| +|2\rangle\langle 1| , \,\,\,\,
  \hat S_{w}=|2\rangle\langle 3| +|3\rangle\langle 2| 
\nonumber\\
\hat S_{h}&=&  |1\rangle\langle 3| +|3\rangle\langle 1|.
\eea
A decohering bath is not included in this example;
as we discuss below in Sec. \ref{sec-res}, in the weak coupling limit 
the system's dynamics is naturally fully secular, with decoupled populations and coherences.

\begin{figure}[htpb]
\vspace{3mm}
\includegraphics[width=8cm]{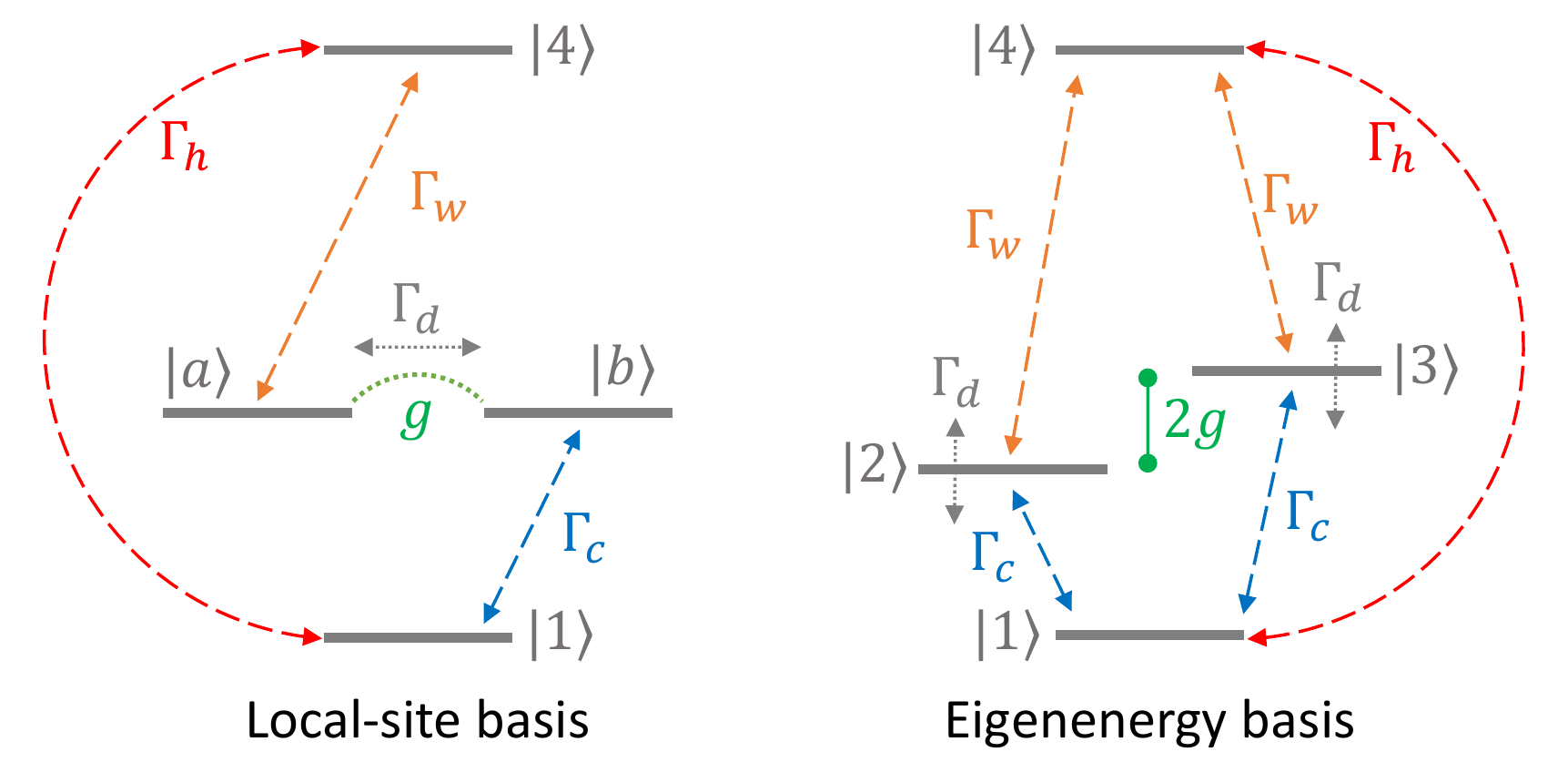}
\caption{
Diagram of the four-level QAR in the local-site basis (left) and energy basis (right).
The red, blue and orange arrows represent energy exchange processes with the hot, cold and work heat baths.
The gray arrow shows the effect of the decoherence bath.
}
\label{fig:4lQAR}
\end{figure}

\subsection{Four-level QAR}

The working medium of the 4lQAR comprises four quantum levels.
It further includes three heat baths ($h$, $c$, $w$), as well as a decohering bath $d$, see Fig. \ref{fig:4lQAR}.
In the site basis, the system's Hamiltonian is
\bea
\hat H_s&=&\epsilon_1|1\rangle\langle 1| +
\theta_c \left(  |a\rangle \langle a|+  |b\rangle \langle b|   \right) +  \theta_h  |4\rangle \langle 4|
\nonumber\\
&+& g\left(|a\rangle \langle b| +|b\rangle \langle a|\right) .
\eea
We again set the reference energy at $\epsilon_1=0$.
The system's operators $\hat S_{\alpha}$ are
\bea
\hat S_{c}&=& |1\rangle\langle b| +|b\rangle\langle 1| , \,\,\,\,
  \hat S_{w}=|a\rangle\langle 4| +|4\rangle\langle a| 
\nonumber\\
\hat S_{h}&=&  |1\rangle\langle 4| +|4\rangle\langle 1|,\,\,\,
\hat S_d= |a\rangle \langle b|+|b\rangle \langle a|.  
\eea
After diagonalization we receive
\bea
\hat H_s=(\theta_c -g) \ |2\rangle \langle 2|
+(\theta_c +g) \ |3\rangle \langle 3|
+\theta_h  |4\rangle \langle 4|,
\eea
and
\bea
\hat S_{c}&=& \frac{1}{\sqrt{2}} \left(|1\rangle\langle 3| -|1\rangle\langle 2| +h.c. \right) , \,\,\,\,
\nonumber\\
 \hat S_{w}&=&\frac{1}{\sqrt{2}} \left( |2\rangle\langle 4| +|3\rangle\langle 4|  +h.c. \right) 
\nonumber\\
\hat S_{h}&=& |1\rangle\langle 4| +|4\rangle\langle 1|, \,\,\,\
\hat S_{d}=  |3\rangle\langle 3| -|2\rangle\langle 2|.
\label{eq:S4l}
\eea
This model is obviously quite complex. Nevertheless, its cooling performance can be explained 
by studying energy transfer in the VETS model and the cooling behavior of the 3lQAR.
A similar model, lacking the dephasing bath, has been explored in the past to address questions of 
heat leaks and endoreversibility of heat engines \cite{jose15}. Nevertheless,
this study was limited to the secular limit, where populations and coherences are decoupled. 
In Sec. \ref{sec-res} we extend this analysis to cover
device operation under the combined effects of internal coupling $g$ and coherence effects, 
while highlighting methodological issues relevant to this setup.


\section{Method and Results}
\label{sec-res}

We study the system's dynamics using the Redfield equation, a
projection operator method \cite{Nitzan}. This method is derived based
on several assumptions: (i) The quantum system weakly couples to its environment, thus
the interaction with the different baths is captured within a second order perturbation theory.
(ii) The reservoirs are Markovian, and (iii) the initial condition is assumed to be system-bath factorized, 
with each bath prepared in a canonical thermal state according to its temperature.
It is important to note that the Redfield equation in general couples coherences, which are the 
off diagonal elements of the reduced density matrix (RDM) $\sigma$, to populations, 
the diagonal elements of $\sigma$.
The Redfield equation is commonly implemented in the system energy basis, though site-basis
implementations were also tested. When the inter-site coupling $g$ becomes large,
such a local quantum master equation provides an incorrect long time solution, 
e.g. showing deviations from the thermal equilibrium state 
\cite{Adesso_2017,Brunner_2017,Michel_2007,Kosloff_2014,Kulkarni16}. 
For a discussion in the context of charge transfer, see for example Refs. 
\cite{Silbey,SegalET00}.  

In the energy basis, the time evolution of $\sigma$ obeys 
the following equation of motion \cite{Nitzan} in the Schr\"odinger representation ($\hbar\equiv 1$),
%
\begin{widetext}
\bea
\dot{\sigma}_{mn}(t)&=&-i\omega_{mn}\sigma_{mn}(t)
-\sum_{\alpha}\sum_{j,k}\bigg[R^\alpha_{mj,jk}(\omega_{kj})\sigma_{kn}(t)+R^{\alpha*}_{nk,kj}(\omega_{jk})\sigma_{mj}(t)
\nonumber\\
&-&[R^\alpha_{kn,mj}(\omega_{jm})+R^{\alpha *}_{jm,nk}(\omega_{kn})]\sigma_{jk}(t)\bigg].
\label{eq:redf}
\eea
\end{widetext}
Here, $\omega_{mn}=\epsilon_m-\epsilon_n$ are frequencies of the system. 
The reservoirs act in an additive manner, a direct outcome of the weak system-bath coupling approximation.
The dissipation terms in Eq. (\ref{eq:redf}) are
\bea
R_{mn,jk} ^{\alpha} (\omega)&=& S_{mn}^{\alpha}  S_{jk}^{\alpha}  \int_0^{\infty} d\tau e^{i\omega \tau}\langle \hat B_{\alpha}(\tau)\hat B_{\alpha}\rangle
\nonumber\\
&=& S_{mn}^{\alpha}  S_{jk}^{\alpha}
\left(\Gamma_{\alpha} (\omega)+i \Delta_{\alpha}(\omega)\right),
\label{eq:Rdef}
\eea     
with matrix elements $S^{\alpha}_{mn}=\langle m|\hat S_\alpha|n\rangle$.
The average is performed with respect to the canonical thermal state (initial condition) of the $\alpha$ bath.
While these $R$ terms, in general, are complex numbers,
in what follows we neglect the imaginary part $\Delta$ since it essentially corresponds to energy shifts.
For harmonic baths and bilinear coupling the real part of the dissipator is
\bea
\Gamma_{\alpha}(\omega)= \begin{cases}
\frac{1}{2}J_\alpha(\omega)n_\alpha(\omega) & \omega<0\\
\frac{1}{2}J_\alpha(\omega)[n_\alpha(\omega)+1] & \omega>0.
\end{cases}
\eea
Here, $J_{\alpha}(\omega)=2\pi\sum_{q}|\lambda_{q,\alpha}|^2\delta(\omega_{q,\alpha}-\omega)$ is the spectral density function of the $\alpha$ bath.
We assume an ohmic function, $J_{\alpha}(\omega)=\gamma_{\alpha}\omega e^{-|\omega|/\omega_c}$;
$\gamma_{\alpha}$ is a dimensionless interaction parameter.
The cutoff frequency $\omega_c$ is the largest energy scale in the problem.
$n_{\alpha}(\omega)$ is the Bose-Einstein occupation function
characterized by the temperature of the $\alpha$ bath, $T_{\alpha}=1/\beta_{\alpha}$ ($k_B\equiv1$).
Note that the temperature of the decohering bath $d$ dictates the magnitude of the decoherence rate, 
yet the $d$ bath does not exchange energy with the system.

While the Redfield equation in general couples the dynamics of populations and coherences, 
one may invoke the so-called secular approximation and decouple these elements. 
This approximation can be justified if the natural (coherent) timescale of the system 
is short relative to its relaxation time---induced by the thermal baths. Furthermore, 
the equations can be effectively ``secularized'' if there is strong decoherence process, $\gamma_d\gg \gamma_{h,c,w}$.

In what follows we study the dynamics of the three different models, the VETS, 3lQAR, and 4lQAR
by exercising the Redfield equation in the energy basis (diagonal $\hat H_s$). 
We further compare results to the secularized behavior (whether justified or not),
so as to expose the role of coherences.
Appendix A includes the full and secularized Redfield equations for the VETS model, and a discussion over their relative applicability.
It should be noted that the Redfield equation is secular by construction for the 3lQAR.

To calculate the energy current we  organize the Redfield equation as
\bea
\dot \sigma= -i[\hat H_s, \sigma(t)] +\sum_{\alpha} \mathcal{D}_{\alpha}\sigma(t),
\eea 
with the dissipators $\mathcal{D}_{\alpha}$.
In the long time limit, $\dot \sigma=0$, we obtain the steady state solution  $\sigma_{ss}$.
Considering the change of the system's energy,
$d\langle \hat H_s\rangle/dt = {\rm Tr}[\hat H_s\dot \sigma]$
(recall that the Hamiltonian does not depend on time) this energy exchange corresponds to heat flow.
With that, we get the standard formula for the (steady state) rate of heat exchange 
with the $\alpha$ reservoir,
\bea
J_{q}^{\alpha}= {\rm Tr}[\hat H_s \mathcal{D}_{\alpha}\sigma_{ss}].
\label{eq:Jq}
\eea
Below we solve equation (\ref{eq:redf}) in the long time limit and find $\sigma_{ss}$. We then calculate
the heat current from Eq. (\ref{eq:Jq}).

\begin{widetext}
\hspace{-35mm}
\begin{figure}[htpb]
{\includegraphics[width=18cm]{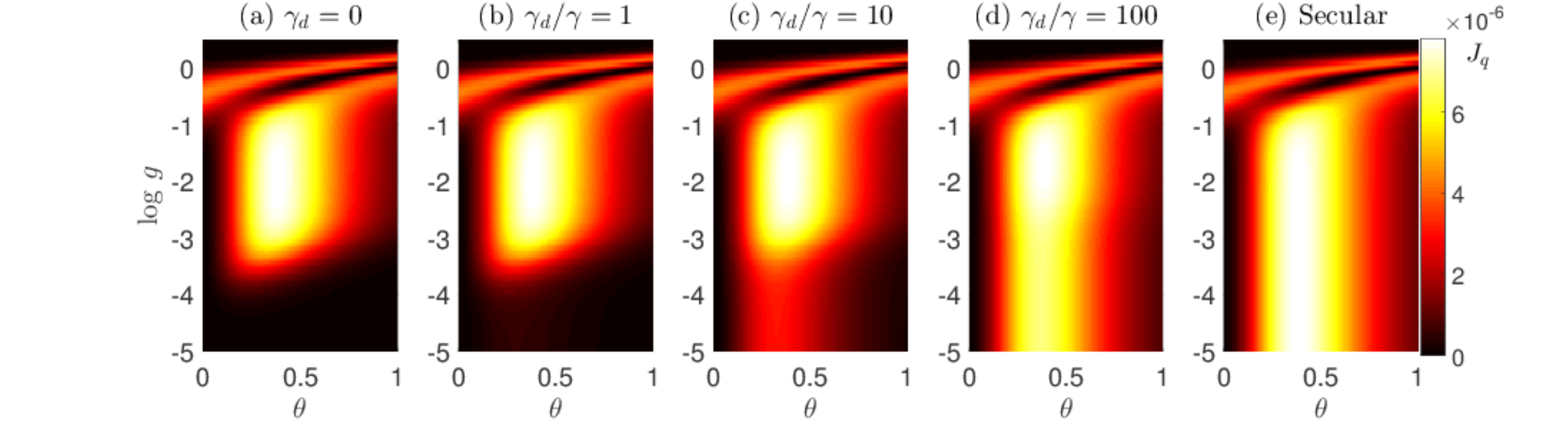}} 
\caption{Heat current in the VETS model.
(a)-(d) Solution of the Redfield equation with an increasing decoherence strength between the excited states.
(e) When decoherence is sufficiently strong, we retrieve the secular behavior.
$\gamma\equiv\gamma_{h,c}=0.002$, $\omega_c=50$, $T_h=0.15$, $T_c=0.1$, $T_d=0.12$.
}
\label{fig:MA1}
\end{figure}
\end{widetext}


\begin{figure}[htpb]
\vspace{5mm}
\hspace{-15mm}
{\includegraphics[width=10cm]{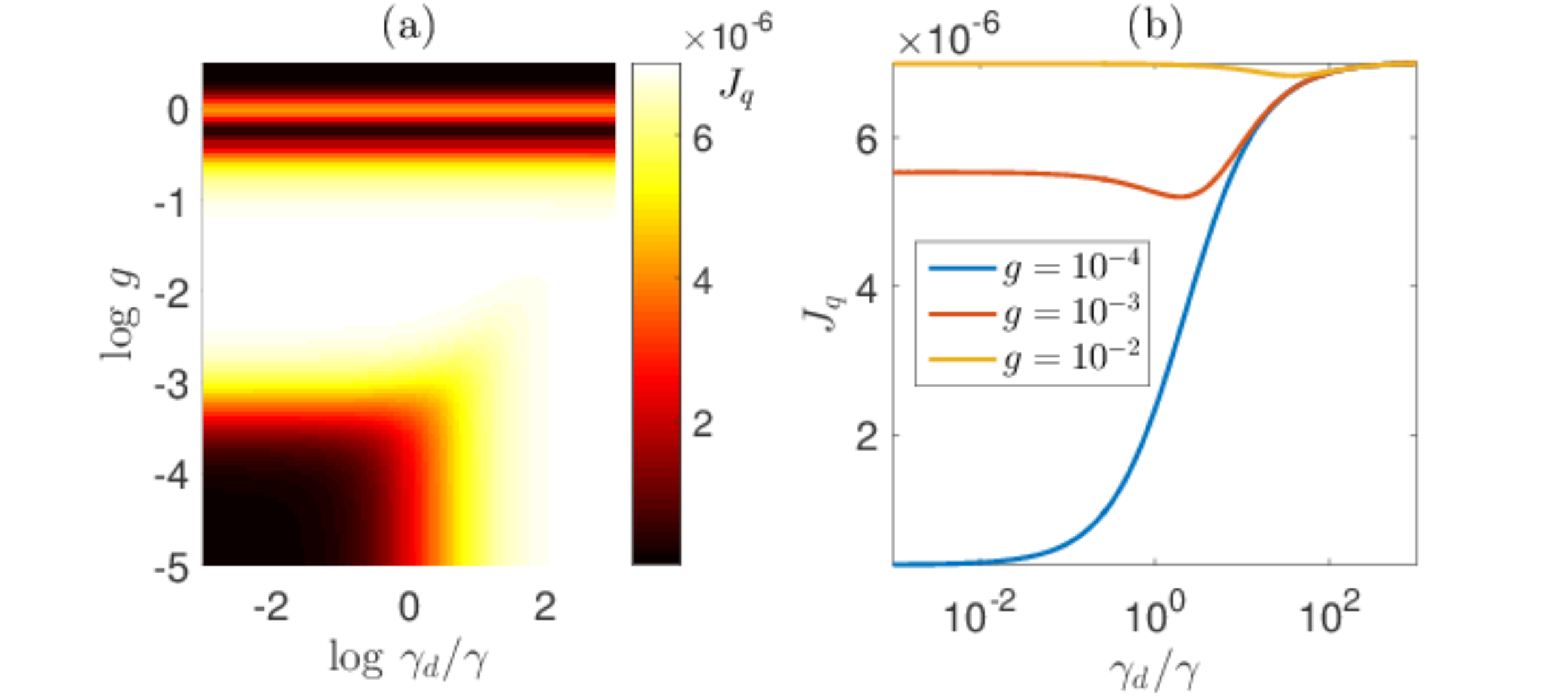}} 
\caption{
(a) Contour plot of current in the VETS model as a function of the intersite coupling 
$g$ and decoherence strength $\gamma_d$.
(b) Demonstration of a non-monotonic $J_q$ vs. $\gamma_d$ behavior.
$\gamma\equiv \gamma_{h,c}=0.002$, $\omega_c=50$, $T_h=0.15$, $T_c=0.1$, $T_d=0.12$, $\theta=0.5$.
}
\label{fig:MA4}
\end{figure}

\subsection{Energy flow in the VETS model}

The Redfield equation of the VETS model is presented in Appendix A and
analytic results for the energy current are derived under the secular approximation.
Fig. \ref{fig:MA1} displays the energy current, from the hot bath to the cold one,
as a function of the energy gap $\theta$ and the coherent coupling $g$.
We further vary the decoherence rate $\gamma_d$ throughout panels (a)-(e). 
The following observations can be made:
(i) In the absence of a decohering bath the heat current approaches zero when $g$ becomes very small,
which is the expected result.
(ii) As we turn on decoherence, the two effective qubits (the pairs 1-2 and 1-3) 
begin to disentangle from each other. Energy then flows between the two heat baths even 
at  small $g$. The large $\gamma_d$ limit
in fact corresponds to the secular behavior, as we show in panel (e).
(iii) The energy current is highly non-monotonic as a function of $g$.
(iv) In our parameters, secularization takes place once $\gamma_d>\gamma_{h,c}$.
(v) The behavior of the energy current as a function of the spacing $\theta$ is intuitive, 
and it follows trends observed for the spin-boson model \cite{nazim}:
When the gap is small,
the energy current increases with $\theta$ since more energy is transferred per quantum.
However, once we further increase the gap, $\theta\gtrsim1/\beta_{\alpha}$,
thermal occupation at relevant-resonant frequencies
is suppressed, and the heat current rapidly drops. 

When coherences are neglected, we are able to reach a simple
closed-form expression for the energy current, see Appendix A for details.
In the asymptotic limit of $g\rightarrow 0$ we get  
\bea
J_q=
\gamma\theta^2\frac{n_h(\theta)-n_c(\theta)}{3n_c(\theta)+3n_h(\theta)+2},
\label{eq:M1Jq}
\eea
where  $\gamma\equiv \gamma_{h,c}$, and $\gamma_d\gg\gamma$.
The current is linear in the coupling parameter $\gamma$, as expected from a weak coupling system-bath description.
The product of the transferred quantum, which is equal to the transition frequency, by the (ohmic) spectral function leads to the quadratic dependence with $\theta$.
Most notably, the current does not vanish at vanishing coupling $g$. 
Without the explicit inclusion of $\gamma_d$, the nonvanishing current $J_q(g\to 0)\neq 0$ 
simply illustrates the breakdown of the secular approximation  at small $g$ (panel (e));
in the local-site basis, $J_q(g\to 0) \to 0$. 
Nevertheless, once  $\gamma_d$ is included this finite current is physically attributed 
to the action of the decohering bath. It couples the $|a\rangle$ and $|b\rangle$ states in the site basis, allowing for energy flow,
see Eq. (\ref{eq:HsNEVS}).

In Fig. \ref{fig:MA4} we display the behavior of the current as a function of $g$ and $\gamma_d$.
Generally, the current increases as we destroy coherences in the system.
Nevertheless, we reveal an intriguing regime, with $J_q$ dropping as we increase $\gamma_d$.
Combining internal couplings with decoherence can therefore impact
energy transfer in a highly non-trivial manner.

\begin{figure}[htpb]
{\includegraphics[width=6cm]{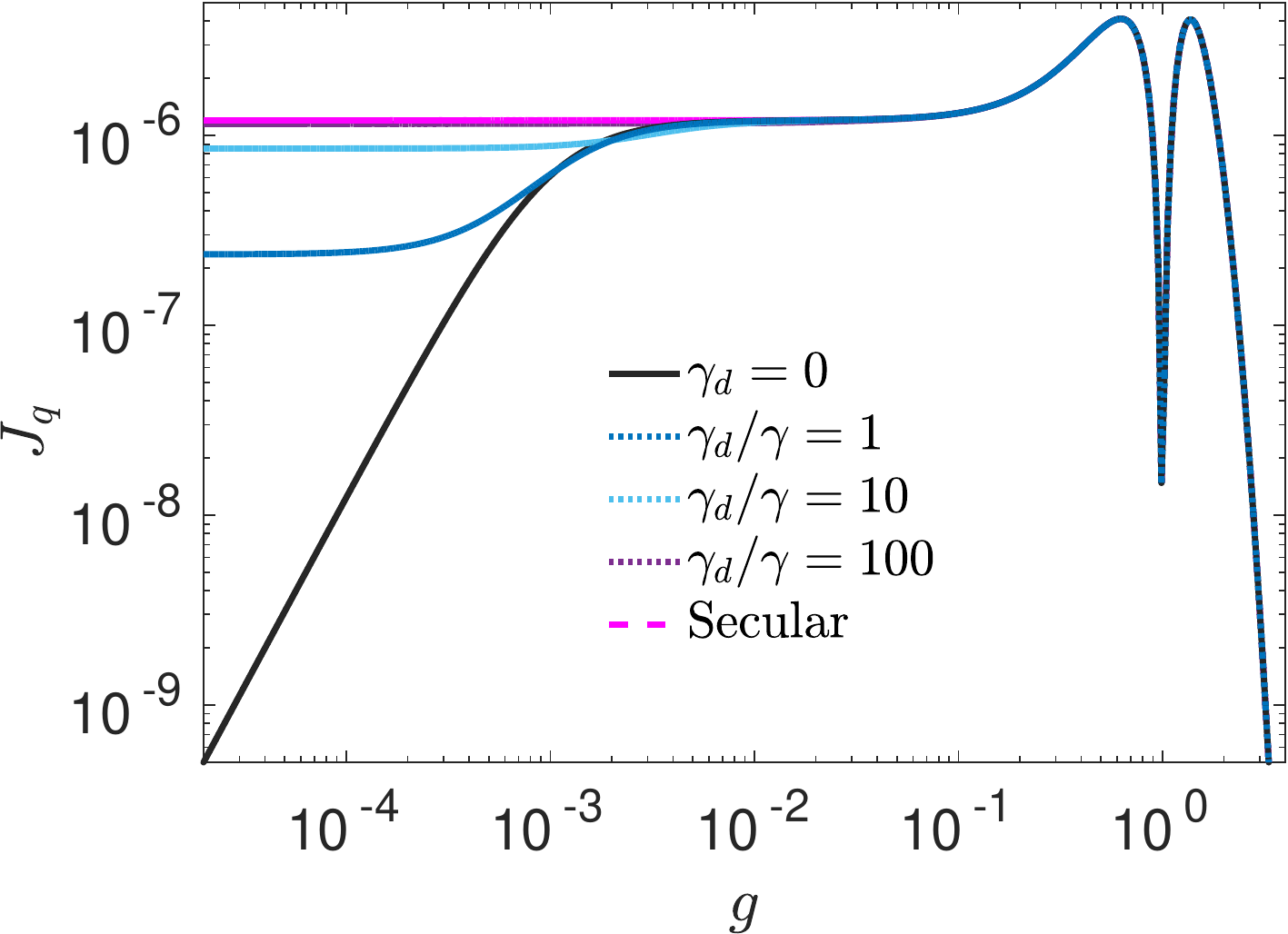}} 
\caption{Heat current in the VETS model as a function of the intersite coupling $g$.
Parameters are the same as in Fig. \ref{fig:MA1}.}
\label{fig:MA2}
\end{figure}

\begin{figure}[htpb]
{ \hspace{1cm}
\includegraphics[width=10cm]{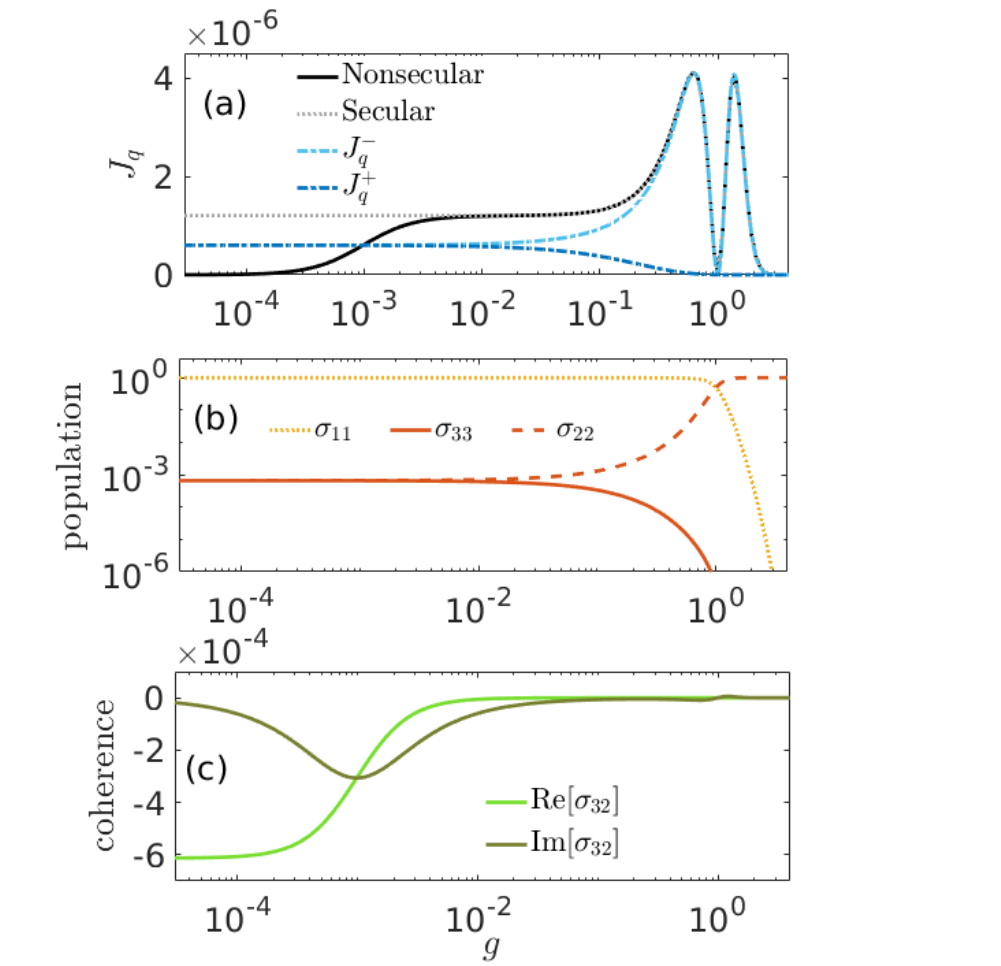}} 
\caption{(a) Current, (b) population and (c) coherences in the VETS model at $\gamma_d=0$.
Parameters are the same as in Fig. \ref{fig:MA1}.}
\label{fig:MA3}
\end{figure}

The behavior of the current as a function of $g$ is quite complex, and we explain it in Figs. 
\ref{fig:MA2}-\ref{fig:MA3}. We further display in Fig. \ref{fig:MA3} 
the currents $J_q^{\pm}$, calculated separately for each effective spin
(of frequency $\theta_{\pm}=\theta\pm g$). 
Relevant energy scales in the problems are $\gamma\theta$, which dictates the hot and cold baths-induced 
decay rates,  $\gamma_d\theta$, which determines the decoherence rate,
the coherent coupling $g$,  the bare energy $\theta$, the temperatures and the reservoir' cutoff frequency,
which is assumed large.
In our simulations,  $\theta\gamma\ll1$.
Focusing on the $\gamma_d=0$ case (black line) in Fig. \ref{fig:MA2}, 
we identify several regimes.

(i) Ultra small coupling, $g\ll \gamma\theta$. 
In this regime, the heat current follows a quadratic scaling $J_q\propto g^2$. 
Fig. \ref{fig:MA3} clearly demonstrates that steady state coherences in the 
system are responsible for the low-$g$
behavior, which deviates from the secular result.
At ultra small $g$, the population of levels 2 and 3 is essentially equal, 
and the magnitude of the coherence is significant.
(ii) Small coupling regime $g\sim\gamma\theta$.  
The current grows with $g$ and then it saturates.
In parallel, the coherence between the excited states drops to zero, 
and the populations of levels 2 and 3 begin to deviate. 
In this regime,  $(\theta_+-\theta_-)/\theta$ grows, eventually  justifying the
secular approximation.
(iii) Intermediate coupling, $\gamma\theta<g <\theta$. 
The heat current quickly rises---then drops.
In this regime, the states 2 and 3 are clearly separated,
thus the role of coherence/decoherence is  inconsequential.
Furthermore, while $\theta_+$ is too large to support energy flow, as it exceeds the thermal energy,
heat can be readily transferred between the two baths
through the $\theta_-$ spin. 
Nevertheless, once $\theta_-$ diminishes, $j_{q}^-\to 0$, 
and the total current is suppressed. 
(iv) Strong coupling, $\theta<g<2\theta$. In this regime the V picture does not hold any longer, since the
eigenstate $|2\rangle$ lies below  $|1\rangle$. 
Nevertheless, we can study energy flow in the model. We find that
as we continue to increase $g$, the total energy current grows, and it is 
transferred through the $\theta_-$ spin.
(v) Ultra strong coupling $g \gg  \theta, 1/\beta_{\alpha}$. 
The current quickly drops with $g$ since both transitions, $\theta_{\pm}$, are  too large to support heat flow,
with the spin  gaps exceeding the thermal energy.
We confirmed that these trends are general, showing up with other choices of $\theta$ and $\beta_{\alpha}$.



The V-type model examined here transfers thermal energy between two heat baths, 
but it does not operate as a machine since it is missing an additional work source.
However, an engine or a cooler can be realized with a V-type model
by e.g. periodically driving the transition frequency $\theta$. 
Such a driven scenario, yet with a perfect degeneracy of excited states 
in the energy basis, was examined in Ref. \cite{Kurizki_2015}. 

\subsection{Four-level QAR with coherences}

Equipped with our understanding of energy flow in the VETS, we are ready to examine 
the operation of quantum absorption refrigerators. 
In Appendix B, we first review the behavior of the 3lQAR model. As we show below,
this model can assist in rationalizing the behavior of the 4lQAR.

We emphasize that in this work we are mostly interested in understanding the cooling performance of the 4lQAR
at small $g$, when internal coherences survive and impact its behavior.
Once $g\sim \gamma\theta$, with the internal couplings being comparable to the relaxation rate constants
to the heat baths, the system's coherences are lost and the behavior becomes
secular.
This incoherent-secular situation was nicely explored in Ref. \cite{jose15},
where different mechanisms responsible for
irreversibility in QARs were classified: internal dissipation, which corresponds to the
competition between cooling pathways that are
optimized at different parameters, and heat leaks, a direct heat exchange between the work
and cold baths.

{\it Cooling Current.}
Based on the Redfield equation, we simulate the system's dynamics 
and the steady state energy exchange with the different baths.
The cooling current of the 4lQAR is presented in Fig. \ref{fig:MC1}, 
where we display only the cooling region, $J_q^c>0$.
Overall, we observe trends similar to the VETS.
First, in panel (a) we show that 
in the absence of decoherence processes, $\gamma_d=0$, 
energy transfer ceases between the states
$|a\rangle$ and $|b\rangle$ at small $g$, and the device cannot function continuously. 
As we increase the decoherence rate constant in panels (b)-(d), 
we gradually recapture the secular behavior 
as depicted in panel (e), with a large cooling current persisting even 
at vanishing $g$. 
The cooling behavior in the 4lQAR is thus analogous to the transport characteristics 
of the VETS model.

When comparing the cooling current of the 4lQAR to the 3lQAR, one can show analytically that 
in the secular, low $g$ limit (when the 4lQAR cooling current is observed to be maximal), 
the 3lQAR outperforms 
the 4lQAR for any choice of parameters. 
One can observe this by simple division of Eqs. (\ref{eq:Jq_3lQAR}) and (\ref{eq:CJqsmall}). 

We further confirmed (not shown) that the behavior of the cooling current as a 
function of $g$ and $\gamma_d$ closely 
follows trends observed in Fig. \ref{fig:MA4}. 
This correspondence includes the small domain of non-monotonicity, with
the cooling current slightly decreasing with an increase of $\gamma_d$ at a 
particular range of $g$. 

\begin{widetext}
\hspace{-35mm}
\begin{figure}[htpb]
{\includegraphics[width=18cm]{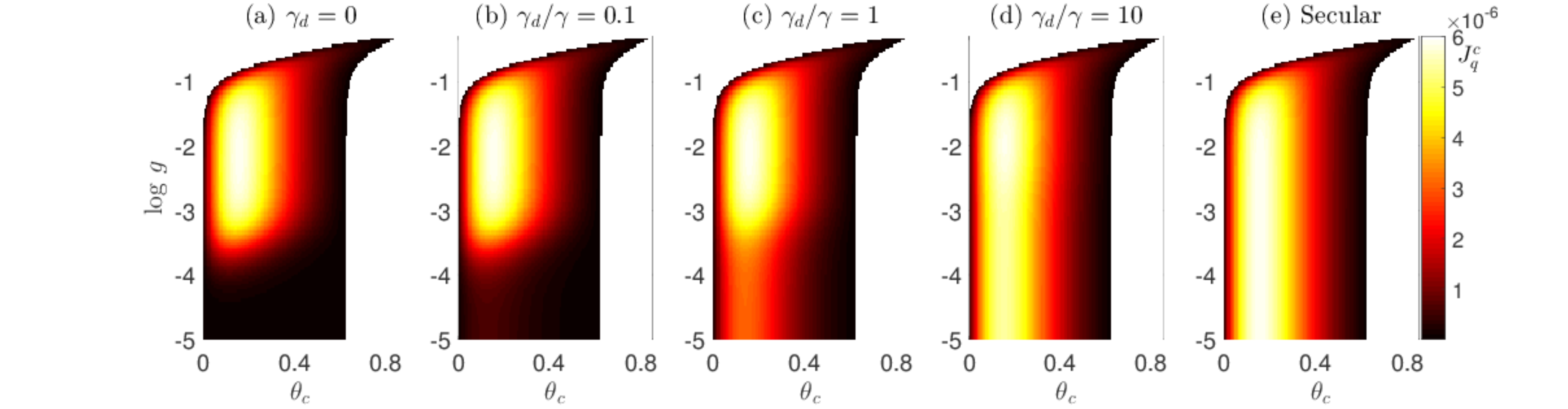}} 
\caption{
Cooling current in the 4lQAR.
(a)-(d) Solution of the Redfield equation with an increasing decoherence rate between the intermediate states.
(e) When $\gamma_d$ is sufficiently strong, we retrieve the secular behavior.
$\gamma\equiv\gamma_{c,h,w}=0.002$, $\omega_c=50$, $T_w=1$, $T_h=0.15$, $T_c=0.1$, $T_d=0.12$, $\theta_h=1$.
}
\label{fig:MC1}
\end{figure}
\end{widetext}

\begin{figure}[htpb]
{\includegraphics[width=7cm]{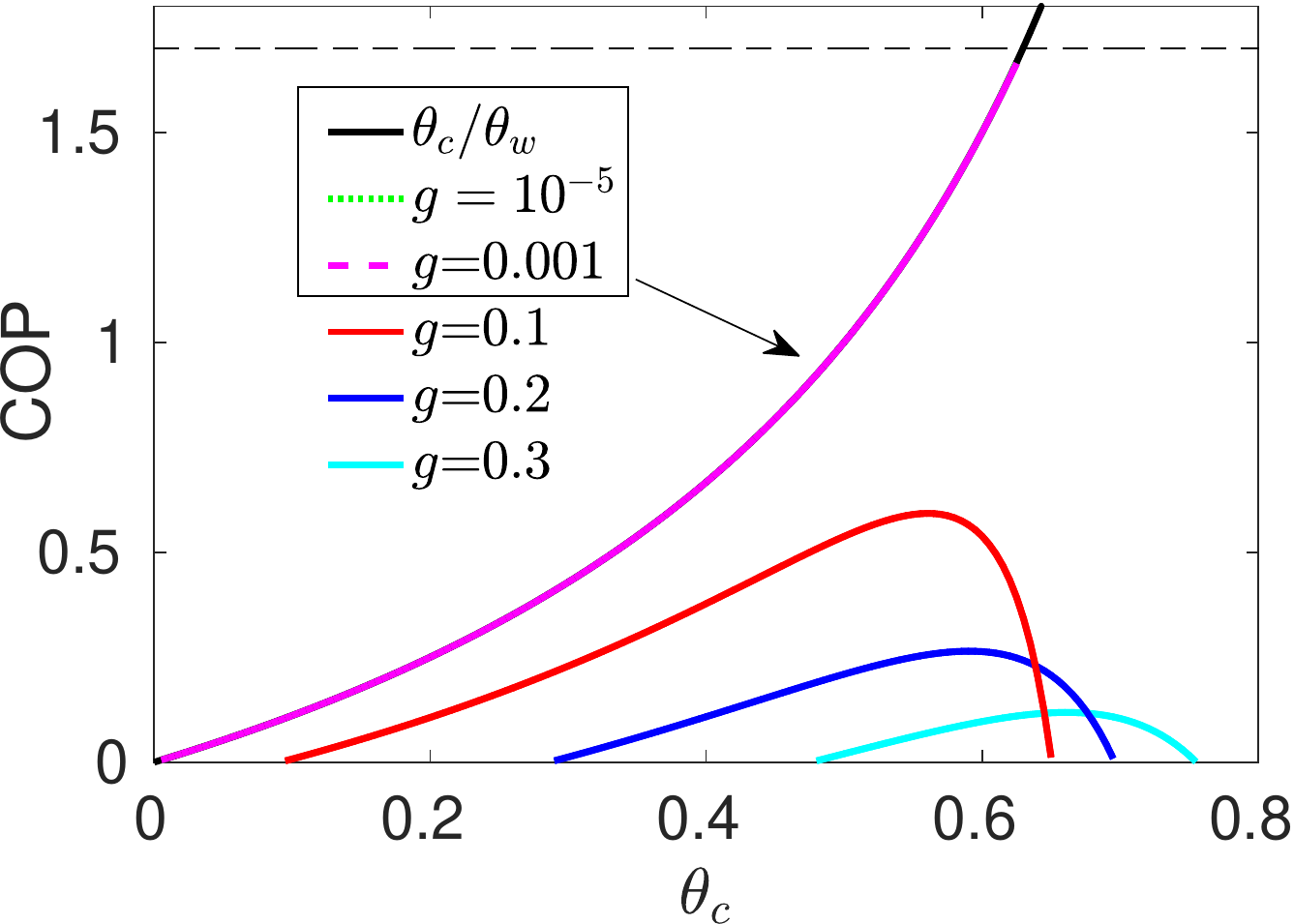}} 
\caption{
COP of the 4lQAR for $\gamma_d=0$ and $\gamma_d/\gamma=100$ (overlapping).
Parameters are the same as in Fig. \ref{fig:MC1}.
The dashed line corresponds to the Carnot bound for a cooling machine, $\eta_C=\frac{\beta_h-\beta_w}{\beta_c-\beta_h}$, which equals 1.7 in our parameters.
}
\label{fig:eff1}
\end{figure}

\begin{figure}[htpb]
{\includegraphics[width=9cm]{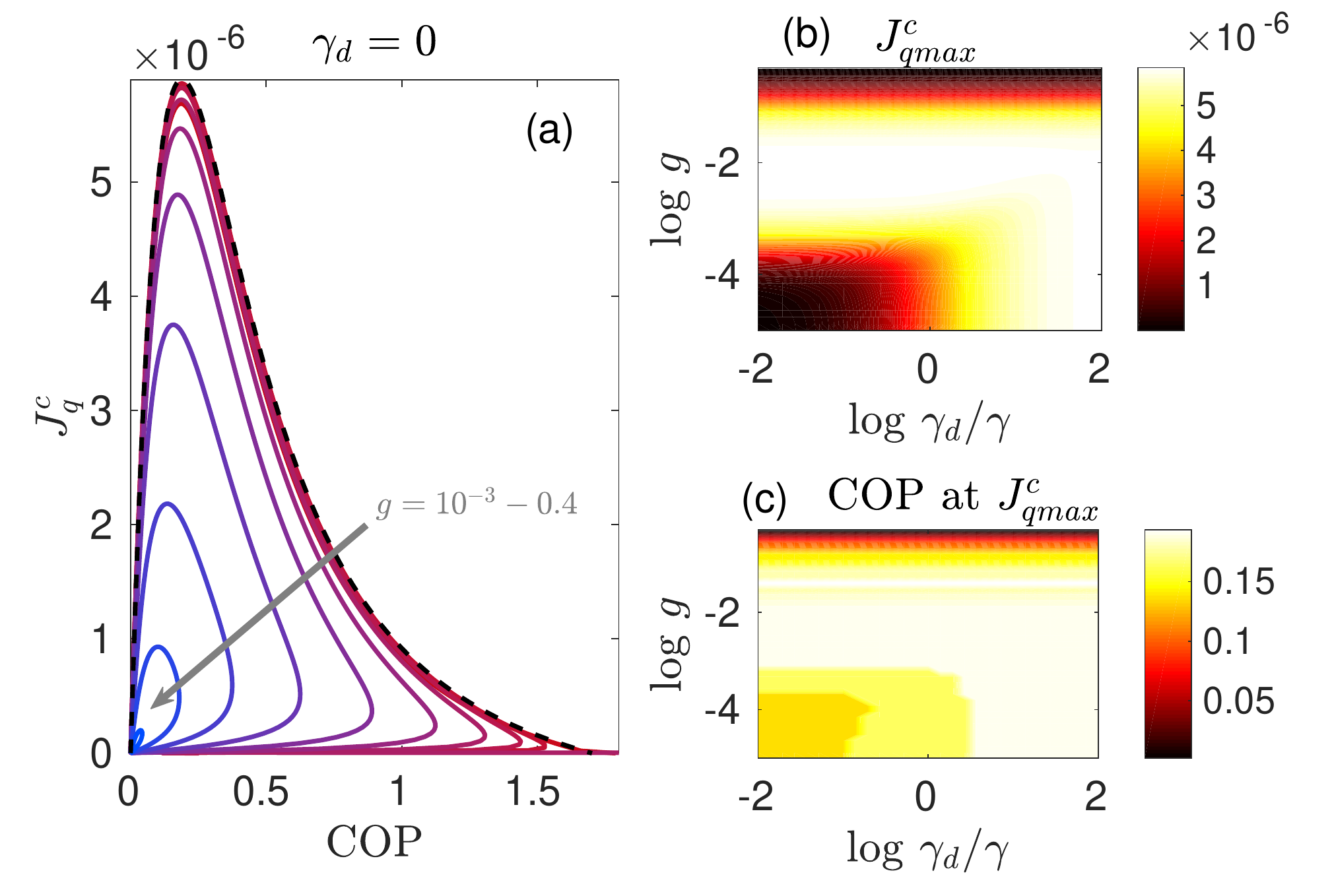}} 
\caption{
(a) COP vs. cooling power for the 4lQAR at $\gamma_d=0$ while increasing $g$ from $10^{-3}$ to 0.4 
as we move inward.
The black dashed line corresponds to the secular limit at vanishing $g$, 
showing an endoreversible operation.  
To generate these plots, the frequency $\theta_c$ was varied within the cooling window. 
(b) Maximum cooling current and (c) COP at that value,
optimized at every point with respect to $\theta_c$.
Parameters are the same as in Fig. \ref{fig:MC1}.
}
\label{fig:eff2}
\end{figure}

{\it Cooling window.}
In Appendices B and C, we derive closed-form expressions for the cooling current and
cooling condition for the 3lQAR and 4lQAR models respectively---under the secular approximation.
In the limit of small $g$ and large $\gamma_d$, the cooling window for the 4lQAR model is
\bea
\frac{\theta_c}{\theta_h} \leq \frac{\beta_h-\beta_w}{\beta_c-\beta_w},
\label{eq:cond_i}
\eea
which is identical to the behavior of the 3lQAR. In the parameters of Fig. \ref{fig:MC1},
this ``secular" cooling window is defined by $\theta_c<0.63$.
In contrast, when $g$ approaches level spacing, $g\sim\theta_c$,
the cooling window of the 4lQAR extends
beyond the $\theta_c=0.63$ bound,  as can be seen by the tail of the cooling window in Fig. \ref{fig:MC1},
developing at large $g$ and large $\theta_c$.

{\it Coefficient of cooling performance.}
We define the coefficient of cooling performance (COP) as $\eta\equiv J_{q}^c/J_{q}^w$, 
and display it in Fig. \ref{fig:eff1}.
In what follows we sometimes refer to the cooling COP as ``efficiency", though it can exceed unity.
Unlike the cooling current, we find from simulations 
that the COP {\it does not depend on $\gamma_d$}. Furthermore, it excellently follows the 
$\eta=\theta_c/\theta_w$ curve, as long as $g$ is small and $\theta_{c,w}$ are maintained as 
a meaningful representation of the spacing.
When $g=0.1-0.3$ the different cooling pathways, through $\theta_c\pm g$,
are differentiated. The behavior in this regime agrees with Ref. \cite{jose15}

In our model, coherences play an important role in dictating the 
magnitude of the cooling current.
However, the cooling COP largely is independent of internal coherences.
In contrast, it is obvious that the efficiency at maximal cooling power  
depends on the internal coupling $g$ and the decoherence rate constant, 
as we show in Fig. \ref{fig:eff2}.
Our results here agree with \cite{jose15}. As $g$ increases,
the endoreversibility of the engine is 
lost due to internal dissipation and heat leaks;
the refrigerator performance suffers significantly in both power and efficiency, 
until we exit the cooling window entirely.

In panels (b) and (c) of Fig. \ref{fig:eff2} we examine the role of decoherence on the performance. 
Panel (b) illustrates the dramatic damping of the current at low $\gamma_d$  and $g$ 
(optimized with respect to $\theta_c$), as observed previously in Fig. \ref{fig:MC1}.
Panel (c) displays the COP at maximal cooling current. Since at strong decoherence the cooling 
behavior of the 4lQAR corresponds to that of the 3lQAR, 
under the ohmic dissipation 
the COP at maximal power should be tightly upper bounded by $\eta_C/2$ \cite{joseSR, Anqi}, 
which is about 0.8 in our parameters. 
Obviously, our parameters are not optimized for performance.


%

\subsection{Experimental realizations}

The energy transfer models described in this work are generic, and may be realized in
atomic or molecular systems \cite{QARE}. 
Moreover, one could envision that the states of the system may represent a coarse grained electronic or
excitonic states in nanoscale electronic or amorphous materials.
Energy conversion systems were studied in e.g. Refs.
\cite{NitzanE1,NitzanE2} based on a classical network picture using a kinetic scheme.
Building on this graph method, which is purely classical, we can imagine a network that
combines both classical (kinetic) and quantum (coherent) transitions.

\begin{figure}[htpb]
\includegraphics[width=8cm]{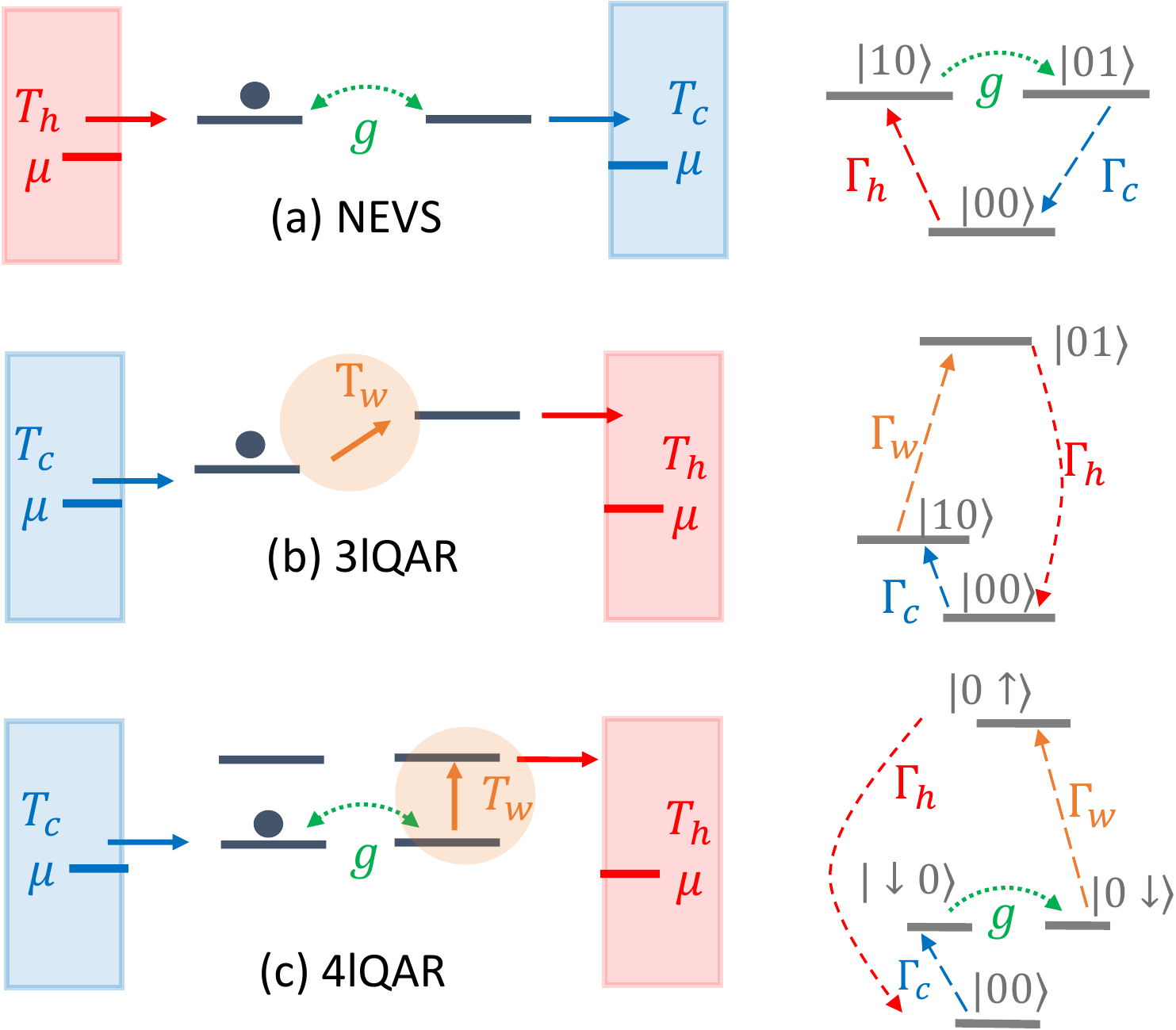}
\caption{
Schemes of nanoscale energy conversion devices that can be captured with our generic models.
(a) The NEVS model can be realized by a conducting junction with two degenerate electronic states.
(b) The 3lQAR may correspond to a non-degenerate electronic junction, with the work bath
responsible for internal transitions.
(c) The 4lQAR can represent a photovoltaic cell. Heat absorbed from the work reservoir 
is used to transfer carriers---thus energy---from the cold metal to the hot metal.
The red, blue and orange arrows represent energy exchange
processes with the hot, cold and work heat baths.
The effect of the decoherence bath is omitted for simplicity, but it can be further included.
}
\label{fig:exp}
\end{figure}

The VETS model of Fig. \ref{fig:VETS} could represent for example a degenerate two-site (termed donor-acceptor)
electronic system embedded between two electrodes, see Fig. \ref{fig:exp}(a).
The electrodes are maintained at the same chemical potentials but at different temperatures, $T_h$ and $T_c$.
In this system,  carriers  transfer energy.
The three states in Fig. \ref{fig:VETS} correspond to empty donor and acceptor sites
$|1\rangle  \leftrightarrow |0 0\rangle$, a single electron residing on the donor site
$|a\rangle\leftrightarrow|1 0\rangle$ and a single electron
 placed on the acceptor site $|b\rangle\leftrightarrow|0 1\rangle$.
Charge transfer  (thus energy flow) between the two metals
is driven by the temperature bias, and it proceeds through
the coherent tunneling element $g$. The overall net process can be summarized as follows,
$|00\rangle \xrightarrow{\Gamma_h} |10\rangle \xrightarrow{g} |01\rangle   \xrightarrow{\Gamma_c}
|00\rangle$.
Alternatively, one could realize the VETS model with two coherently coupled degenerate qubits 
that mediate energy flow between two heat baths through their excitation and de-excitation,
see e.g. Refs. \cite{Wu11,Tanimura15}.

The 3lQAR utilizes input heat from the work reservoir to extract energy from a cold bath.
As we show in Fig. \ref{fig:exp}(b), the model can be realized with
a two-site electronic system (or two qubits), similarly to the VETS.
However, in the 3lQAR model the donor and  acceptor sites are no longer degenerate,
with the energy of the donor state placed below the acceptor. As well, in this scenario all transitions
are kinetic. The cooling process corresponds to
$|00\rangle \xrightarrow{\Gamma_c} |10\rangle \xrightarrow{\Gamma_w} |01\rangle   \xrightarrow{\Gamma_h}
|00\rangle$.

The 4lQAR can be realized within a photovoltaic cell, with the current flowing between
the two metals being assisted by the absorption of solar energy ($T_w$).
The system comprises two degenerate species, donor and acceptor,
each represented by a ground state and an excited state (HOMO and LUMO, respectively, in the language
of molecular orbitals).
We assume that the donor is only coupled to the left (cold) metal, while the acceptor
is coupled to the right (hot) lead. Furthermore,
the acceptor is excited by the solar energy, 
allowing energy to flow against the temperature difference.
The model is sketched in Fig. \ref{fig:exp}(c).
The four states of the 4lQAR correspond to an empty system $|1\rangle \leftrightarrow |0 0\rangle$,
the donor specie in the HOMO state with an empty acceptor site
 $|b\rangle \leftrightarrow |\downarrow 0\rangle$,
the complementary state with the charge localized on the  acceptor site $|a\rangle \leftrightarrow |0 \downarrow \rangle$,
and the highest energy state,
with the acceptor in the excited (LUMO) state $|4\rangle \leftrightarrow |0 \uparrow \rangle$.
The $|a\rangle $ and $|b\rangle$ states are coherently coupled with a tunneling element $g$.
A cooling process corresponds to the cycle
$|00\rangle \xrightarrow{\Gamma_c} |\downarrow 0\rangle \xrightarrow{g} |0\downarrow\rangle  
 \xrightarrow{\Gamma_w }  |0\uparrow\rangle$  $\xrightarrow{\Gamma_h }  |0 0\rangle$.
Alternatively, the 4lQAR can be realized by three sequential qubits.
Excitations transfer coherently between the first two degenerate qubits, 
and incoherently (assisted by the work bath) between the second and third nondegenerate qubits.

The multi-level models of heat machines explored in this work can therefore represent a variety of
devices, such as photovoltaic cells and thermoelectric junctions,
while making use of both classical-kinetic and quantum-coherent resources.


\section{Summary}
\label{sec-summ}

We studied the cooling performance of a generic multi-level QAR with steady state bath-induced coherences.
The central observations of this work are the following:

(i) Generally, the presence of coherences in our system degrade the cooling performance of the QAR. 
Introducing decoherence essentially ``secularizes" the coherent dynamics, 
by suppressing the off-diagonal elements of 
the reduced density matrix. As a result, at strong decoherence we re-capture the secular dynamics 
for any internal coupling $g$.  
One could interpret this observation by realizing that the survival of coherences in our 
setup corresponds to state 
delocalization, which is detrimental to the operation of the QAR.
To optimize the performance of our QAR one needs to minimize leaks and internal dissipation (keep $g$ small), 
while simultaneously minimizing delocalization. 
This conclusion suggests that in certain designs, strong system-bath coupling effects could enhance performance, by promoting localized dissipation.

(ii) Varying the internal coupling $g$ within the system results in rich behavior 
as the QAR crosses between different regimes of operation.
Moreover, the interplay of internal coupling in the system and decoherence rate
reveals a small non-monotonic effect of coherence in the device power.

(iii) The cooling efficiency of the refrigerator, defined as the ratio of
extracted heat from the cold bath over input heat from he work reservoir,
does not depend on the survival of steady state coherences in the system.
In fact, the cooling efficiency of the 4lQAR precisely follows the behavior of 
the incoherent 3lQAR model when the internal coupling, $g$, is small.
In contrast, the efficiency at maximal cooling current is quite sensitive to the magnitude of the
internal couplings and therefore can be affected and manipulated by internal coherences.

(iv) The Redfield equation, despite its known pathologies regarding positivity, 
is well-behaved at steady state across a very wide parameter regime explored in this work, 
and it gives numerical and analytical insight missing in other approaches. 
The global and local secular quantum master equations catastrophically fail outside 
their respective regimes of applicability. This fact is 
not surprising, yet it is important to visualize this failure in the context of QHMs. 
Specifically, we found that when the global secular equation is applied to a fundamentally nonsecular system
it provides erroneous predictions for the cooling current. 
Nevertheless, our simulations demonstrated that it gives correct results for the cooling COP. 

(vi) More generally, our work demonstrates that the operation of 
a composite quantum heat machine 
can be explained from the behavior of its components, 
allowing us to identify key operational principles in energy conversion devices.


Going forward, there are several avenues that should be explored in this area. 
First, an interesting result of this study which deserves future attention is the 
insensitivity of the QAR efficiency to coherence/decoherence, 
in contrast to the quite dramatic effects observed in the current. 
Future work will examine the connection between efficiency and coherences in QHMs. 
While we had assessed the system's quantumness by studying its coherences 
(magnitude of off diagonal elements of the reduced density matrix), 
it is interesting to further examine other measures for quantumness and delocalization in quantum
energy transport systems and QHMs, such as state purity, inverse participation ratio, 
entanglement and discord \cite{Goold,Wu11,Tanimura15}.

While quantum thermodynamical machines have been traditionally analyzed under a strict 
weak-coupling approximation, it is now recognized that to properly characterize the 
performance of quantum engines and refrigerators, and moreover to achieve new functionality,
one must develop methods that are not limited in this respect 
\cite{nazim,Anqi,DavidS,KosloffGil,Cao,Cao_2018,Gernot,Tanimura,Jarzynski,esposito17,Nazir,Celardo,Friedman,Tanimura18}. 
The exploration of quantum coherent effects in strongly coupled heat machines is left for future work.

%
\begin{acknowledgments}
Dvira Segal acknowledges support from an NSERC Discovery Grant and the Canada Research Chair program.
The work of Michael Kilgour was supported by an NSERC Canada Graduate Scholarship-Doctoral (CGS D). 
\end{acknowledgments}

\renewcommand{\theequation}{A\arabic{equation}}
\setcounter{equation}{0}  

\section*{Appendix A: Energy transfer in the V system }
\label{app:NEVS}

We derive here a closed-form expression for the energy current in the VETS model under the secular approximation.
Our starting point is the Redfield equation (\ref{eq:redf}) for the reduced density matrix $\sigma$, 
written in the energy basis (``global Redfield"),
which couples the dynamics of populations and coherences.
For the VETS model (\ref{eq:HNEVSd})-(\ref{eq:SNEVS}), the Redfield equation reduces to
\begin{widetext}
\bea
\dot{\sigma}_{11}(t) &=&
-(R_{12,21}+R^*_{12,21}+R_{13,31}+R^*_{13,31})\sigma_{11}(t)+(R_{21,12}+R^*_{21,12})\sigma_{22}(t)+(R_{31,13}+R^*_{31,13})\sigma_{33}(t)
\nonumber\\
&+&{\bf (R_{31,12}+R^*_{21,13})}\sigma_{23}(t)+({\bf R_{21,13}+R^*_{31,12}})\sigma_{32}(t),
\nonumber\\
\dot{\sigma}_{22}(t)&=&
-(R_{21,12}+R^*_{21,12})\sigma_{22}(t)+(R_{12,21}+R^*_{12,21})\sigma_{11}(t)-{\bf R^*_{21,13}}\sigma_{23(t)}-{\bf R_{21,13}}\sigma_{32}(t),
\nonumber\\
\dot{\sigma}_{33}(t)&=&
-(R_{31,13}+R^*_{31,13})\sigma_{33}(t)+(R_{13,31}+R^*_{13,31})\sigma_{11}(t)- 
{\bf R_{31,12}}\sigma_{23}(t)-{\bf R^*_{31,12}}\sigma_{32}(t),
\nonumber\\
\dot{\sigma}_{23}(t)&=& +2ig \sigma_{23}(t)
-(R_{21,12}+R^*_{31,13})\sigma_{23}(t) +({\bf R_{13,21}+R^*_{12,31}})\sigma_{11}(t)-{\bf R^*_{31,12}}\sigma_{22}(t)-{\bf R_{21,13}}\sigma_{33}(t)
\nonumber\\
&-& ( R_{22,22}^{d}+ R_{33,33}^{d,*}  -R_{33,22}^{d} - R_{22,33}^{d,*}  )\sigma_{23}(t),
\nonumber\\
\dot{\sigma}_{32}(t)&=& -2ig \sigma_{32}(t)
-( R^*_{21,12}+R_{31,13})\sigma_{32}(t) + ({\bf R_{12,31}+R^*_{13,21}})\sigma_{11}(t)-{\bf R_{31,12}} \sigma_{22}(t)-{\bf R^*_{21,13}}\sigma_{33}(t)
\nonumber\\
&-& ( R_{22,22}^{d,*}+ R_{33,33}^{d}  -R_{22,33}^{d} - R_{33,22}^{d,*}  )\sigma_{32}(t).
\nonumber\\
\label{eq:redfV}
\eea
\end{widetext}
For simplicity, we do not explicitly indicate the frequency at which the dissipation elements are calculated,
see Eqs. (\ref{eq:redf})-(\ref{eq:Rdef}).
Note that there are three types of  bath-induced terms in Eq. (\ref{eq:redfV}): 
(i) We put together the rate constants induced by the hot and cold baths,
$R_{mn,jk}(\omega)= \sum_{\alpha=h,c}R_{mn,jk}^{\alpha} (\omega)$.
For example, $R_{12,21}(\omega)= \Gamma_h(\omega) + \Gamma_c(\omega)$ (ignoring for simplicity the imaginary term).
(ii) On top of that, we use bold faced fonts to highlight terms in which the sum $\sum_{\alpha=h,c}R_{mn,jk}^{\alpha} (\omega)$
turns into the difference between the hot and cold 
dissipation rates, such as in $R_{21,13}(\omega)= \Gamma_h(\omega) - \Gamma_c(\omega)$.
The sign difference develops because
the operators $\hat{S_{h}}$  and $\hat{S_{c}}$  have opposite signs 
in front of the transitions $|1\rangle\langle 2|+ h.c.$, see Eq. (\ref{eq:SNEVS}).
(iii) We separately identify contributions from the decohering bath, for example,
$R_{22,22}^d=-R^d_{22,33}=\Gamma_d(0)$. 
Because $n(0)$ diverges, the zero-frequency transitions mediated through the decohering bath 
require some extra effort to compute. 
In general, this term is dependent on the functional form of the spectral density. For an ohmic
bath, we employ L'H\^opital's rule to compute the limit
\begin{equation}
\begin{split}
J(0)n(0)=&\lim_{\omega\to 0}\frac{\gamma\omega e^{-|\omega|/\omega_c}}{e^{\beta\omega}-1}\\
=&\lim_{\omega\to 0}\frac{\gamma e^{-|\omega|/\omega_c}(\omega_c-\omega)}{\omega_c\beta e^{\beta\omega}}=\frac{\gamma}{\beta}.
\end{split}
\end{equation}
Based on Eq. (\ref{eq:redfV}), we can clearly expose the conditions for which a secular approximation is justified:

(i) Eigenbasis coherences only survive in the long-time limit if the 
system is driven out of equilibrium, i.e. there is a temperature bias between the reservoirs. 
If there is no temperature bias, the system will adhere to the Gibbs state defined by the bath temperatures, 
which is completely diagonal. 
This can be proved as follows. When $T_h=T_c$, we add up the equations of motion for the coherences,
 $\dot \sigma_{23}+\dot\sigma_{32}$, and in the long time limit,
we assume zero coherences.
The only surviving terms then (in bold) have a sign flip between the two (equal temperature) baths, 
${\bf R_{ij,kl}(\omega_{lk})}= \Gamma_h(\omega_{lk}) - \Gamma_c(\omega_{lk})$. This contribution is proportional to ($\gamma_h-\gamma_c$),
resulting in a trivial identity  for the population dynamics.
Since the algebraic steady state equation has a unique solution, the trial solution of zero coherences, which does not show contradictions, is valid.

(ii) When $g$ is large compared to the dissipation rate, 
the so-called ``rotating-wave approximation'' becomes well-justified. 
This translates to the timescale of coherence oscillation being short 
compared to other timescales of the system, 
and therefore effectively decoupling coherences from the population dynamics.

(iii) Finally, one can directly ``secularize'' the dynamics by adding strong decoherence effects, $\gamma_d\gg\gamma_{h,c,w}$.
 This contribution exponentially damps the coherence terms, $\sigma_{23},\sigma_{32}$, resulting in a secular dynamics. 

Working under the secular approximation with population and coherence dynamics decoupled,
we find from Eq. (\ref{eq:redfV}) that the 
populations follow a kinetic equation
$ |\dot p\rangle = \sum_{\alpha=h,c}\mathcal{D}_{\alpha} |p\rangle$, 
with the vector $|p\rangle=(p_1,p_2,p_3)^{{\mathsf T}}$ and the dissipator
\begin{equation}
\mathcal{D}_{\alpha}=
\begin{bmatrix}
 -k_{1\to 2}^{\alpha} -k_{1\to 3}^{\alpha} 
& k_{2\to 1}^{\alpha}
& k_{3\to 1}^{\alpha}\\
k_{1\to 2}^{\alpha}
&-k_{2\to 1}^{\alpha}&0\\
k_{1\to 3}^{\alpha}&0&
-k_{3\to 1}^{\alpha}
\end{bmatrix}
\end{equation}
It is useful to define the two relevant frequencies as $\theta_{\pm}=\theta\pm g$, see Fig. \ref{fig:VETS}.
The rate constants combine two Redfield dissipation rates,
for example, $k_{1\to 2}^h=R_{12,21}^h+R_{12,21}^{h,*}$. Explicitly, they are given by
$k_{1\to 2}^{\alpha}=\frac{1}{2}J_{\alpha}(\theta_-) n_{\alpha}(\theta_-)$ and
$k_{1\to 3}^{\alpha}=\frac{1}{2}J_{\alpha}(\theta_+) n_{\alpha}(\theta_+)$,
with the  reversed rates $k_{2\to 1}^{\alpha}$ and $k_{3\to1}^{\alpha}$ determined from detailed balance.
Note, the factor 1/2 within these expressions results from the $1/\sqrt2$ prefactor in front of the eigenbasis 
operators $\hat S_{\alpha}$ in Eq.
(\ref{eq:SNEVS}). 
Here, $n_{\alpha}(\theta_{\pm})=\left( e^{\beta_{\alpha}\theta_{\pm}} -1 \right)^{-1}$ 
is the Bose-Einstein distribution function
and $J_{\alpha}(\theta_{\pm})$ is the spectral density of the $\alpha$ bath.
In the ohmic limit with a large cutoff frequency, $\omega_c \gg \theta, g$, 
\bea
k_{1\to 2}^{\alpha}=\frac{1}{2}\gamma_{\alpha} \theta_- n_{\alpha}^-, \,\,\,
k_{1\to 3}^{\alpha}=\frac{1}{2}\gamma_{\alpha} \theta_+n_{\alpha}^+,
\label{eq:Arates}
\eea
We use the short notation 
$n_{\alpha}^{\pm}\equiv n_{\alpha}(\theta_{\pm})$.
For simplicity, we further assume a symmetric coupling, 
$\gamma=\gamma_{c,h}$. Recall that $\gamma_{h,c,d}$ are dimensionless coupling parameters.

We readily calculate the energy current (e.g. from the hot bath) using Eq. (\ref{eq:Jq}), 
or directly with the approach described in Ref. \cite{QARfluct},
where we construct partial cumulant generating functions.
The energy current from the hot bath includes two contributions from the two transitions (``spins"), 
\bea
J_q&=&
\theta_-\frac{\Big[\left(k_{1\to2}^hk_{2\to 1}^c-k_{1\to 2}^ck_{2\to 1}^h )
(k_{3\to 1}^c+k_{3\to 1}^h\right)\Big]}{\Psi}
\nonumber\\
&+&\theta_{+} \frac{\Big[ \left(k_{1\to3}^hk_{3\to 1}^c-k_{1\to 3}^ck_{3\to 1}^h )
(k_{2\to 1}^c+k_{2\to 1}^h\right)\Big]}{\Psi}.
\label{eq:AJq}
\nonumber\\
\eea
The denominator is 
\bea
\Psi&=&\left(k_{1\to3}^c +k_{1\to3}^{h}\right)
\left(k_{2\to1}^c+k_{2\to1}^h\right) 
\nonumber\\
&+& \left(k_{3\to1}^c+k_{3\to1}^h\right)
\left( k_{2\to1}^h+k_{2\to1}^c +k_{1\to2}^c +k_{1\to2}^h\right).
\nonumber
\eea
Equation (\ref{eq:AJq}) describes the energy current in the NEVS model when the population and coherences dynamics
are decoupled. 
%
The energy current  depends on the inter-site coupling $g$ in a rich manner.
First, clearly, the frequencies $\theta_{\pm}$ are linear functions in $g$. 
More involved behavior arises since the distribution function 
$n_{\alpha}^{\pm}$ further depends on the transition
frequency, thus it is a nonlinear function of $g$.

To the lowest order in $g$, the levels $\theta_{\pm}$ are degenerate thus 
$k^{\alpha}_{1\to 2}=k^{\alpha}_{1\to 3}$. 
As a result, the current (\ref{eq:AJq}) reduces to
\bea
J_q=2\theta \frac{\left[ k_{1\to 2}^hk_{2\to 1}^c - k_{1\to 2}^c k_{2\to 1}^h\right]}
{ 2(k_{1\to 2}^c +k_{1\to 2}^h) + k_{2\to1}^c+k_{2\to 1}^h}.
\eea
%
Using the concrete expressions for the rate constants (\ref{eq:Arates}), we immediately get
\bea
J_q=\gamma\theta^2\frac{n_h(\theta)-n_c(\theta)}{3n_c(\theta)+3n_h(\theta)+2}.
\label{eq:jqsmallg}
\eea 
%
Beyond the lowest order in $g$, corrections are given by a power series,  $g^{2n}$ with $n=1,2,3,...$.
We reiterate that Eq. (\ref{eq:jqsmallg}) is valid when the internal coupling $g$ is small yet the decoherence
constant $\gamma_d$ is large enough such that the secular approximation is well founded.

\renewcommand{\theequation}{B\arabic{equation}}
\setcounter{equation}{0}  

\section*{Appendix B: 3lQAR}
\label{app:3lQAR}

We derive here an expression for the cooling current in the canonical three-level QAR model.
We further discuss the cooling window and the cooler coefficient of performance (termed efficiency here).
In the 3lQAR model, each transition is coupled to a separate thermal bath; internal couplings between
system's states are absent. As a result,
in this model population and coherences naturally decouple, without the need to further invoke the secular
approximation. 
Using the Redfield equation in a manner analogous to that described in Appendix A,
we find that the population dynamics follows a kinetic equation
$ |\dot p\rangle = \sum_{\alpha=c,h,w}\mathcal{D}_{\alpha} |p\rangle$,
with the population vector $|p\rangle=(p_1,p_2,p_3)^{{\mathsf T}}$, and the total dissipator
\begin{equation}
\mathcal{\sum_{\alpha}D_{\alpha}}=
\begin{bmatrix}
 -k_{1\to 2}^{c}  
& k_{2\to 1}^{c}
& k_{3\to 1}^{h}\\
k_{1\to 2}^{c}
&-k_{2\to 1}^{c}  -k_{2\to 3}^{w} & k_{3\to 2}^{w}\\
k_{1\to 3}^{h}&  k_{2\to 3}^{w}&
-k_{3\to 1}^{h}  -k_{3\to 2}^{w}
\end{bmatrix}
\end{equation}
The rate constants for a given transition
are given by the Bose-Einstein function and the spectral density, evaluated at the transition frequency,
$k_{1\to 2}^{c}=J_{c}(\theta_c) n_{c}(\theta_c)$,
$k_{1\to 3}^{h}=J_{h}(\theta_h) n_{h}(\theta_h)$ and
$k_{2\to 3}^{w}=J_{w}(\theta_w) n_{w}(\theta_w)$.
The  reversed rate constants
are determined from detailed balance.
We assume ohmic spectral functions with  large cutoff frequencies.
For simplicity, we further take the couplings to be identical at all contacts, $\gamma=\gamma_{\alpha}$.
Since the Bose-Einstein function is evaluated at the transition frequency,
we can use the short notation $n_{\alpha}\equiv  n_{\alpha}(\theta_{\alpha})$.
%
Following Eq. (\ref{eq:Jq}), the heat current at the $c$ bath is given by
\bea
J_q^c=\theta_c\frac{k_{1\to 2}^c k_{2\to 3}^w k_{3\to 1}^h - k_{1\to 3}^h k_{2\to 1}^c k_{3\to 2}^w}{\Psi},
\label{eq:Jq_3lQAR}
\eea
with the denominator
\bea
\Psi&=&  k_{1\to3}^h k_{2\to1}^c + k_{1\to2}^ck_{2\to3}^w + k_{1\to3}^h k_{2\to3}^w 
\nonumber\\
&+& k_{1\to2}^c k_{3\to1}^h  +  k_{1\to2}^c k_{3\to2}^w + k_{1\to3}^h k_{3\to2}^w 
\nonumber\\
&+& k_{2\to1}^c k_{3\to1}^h  + k_{2\to1}^c k_{3\to2}^w + k_{2\to3}^w k_{3\to1}^h.
\eea
Explicitly, we find that the cooling current is given by
\bea
J_q^c\propto  \gamma\theta_c  \left[ n_cn_w(n_h+1) - n_h(n_c+1)(n_w+1) \right].
\eea
Positive cooling current, i.e. the 
extraction of thermal energy from the cold bath, is achieved when the inequality is satisfied,
\bea
(n_c+1)(n_w+1)n_h < (n_h+1)n_w n_c. 
\eea
One can then readily identify the cooling window,
\bea
\frac{\theta_c}{\theta_h} \leq \frac{\beta_h-\beta_w}{\beta_c-\beta_w}.
\label{eq:cond}
\eea
It can be also shown that the cooling coefficient of performance is
\bea
\eta\equiv J_q^c/J_q^w=\theta_c/\theta_w,
\eea
and that this coefficient is bounded by the Carnot limit,
\bea
\eta_C=\frac{\beta_h-\beta_w}{\beta_c-\beta_h}.
\eea 
For the ohmic model, the efficiency at maximal power is 
tightly bounded by half of the Carnot efficiency \cite{joseSR}.

\renewcommand{\theequation}{C\arabic{equation}}
\setcounter{equation}{0}  

\section*{Appendix C: 4lQAR}
\label{app:4lQAR}

The 4lQAR combines features from both the NEVS and the 3lQAR.
As described in Appendix A, our starting point is the Redfield equation, working as usual in the energy basis.
The structure of the equation would be analogous to (\ref{eq:redfV}), with an additional level $|4\rangle$ 
and the work bath 
$w$.

We can easily calculate the cooling current in the 4lQAR model within the secular assumption.
First, we define four frequencies, $\theta_c^{\pm}=\theta_c\pm g$, $\theta_w^{\pm}=\theta_w\pm g$,
which obey $\theta_h=\theta_c^{\pm} + \theta_w^{\mp}$. 
%
Under the secular approximation,
the populations follow a kinetic equation of motion,
$ |\dot p \rangle = \sum_{\alpha=h,c,w}\mathcal{D}_{\alpha} |p\rangle$, 
with the population vector $|p\rangle=(p_1,p_2,p_3,p_4)^{{\mathsf T}}$ and the total dissipator,
\begin{widetext}
\begin{equation}
\mathcal{\sum_{\alpha}D_{\alpha}}=
\begin{bmatrix}
-k_{1\to 2}^{c} -k_{1\to 3}^{c} -k_{1\to 4}^h & k_{2\to 1}^{c} & k_{3\to 1}^{c} & k_{4\to 1}^h\\
k_{1\to 2}^{c} &-k_{2\to 1}^{c} -k_{2\to 4}^w&0 & k_{4\to 2}^w \\
k_{1\to 3}^{c}&0& -k_{3\to 1}^{c} -k_{3\to 4}^w & k_{4\to 3}^w\\
k_{1\to 4}^h& k_{2\to 4}^w&k_{3\to 4}^w& -k_{4\to 1}^h-k_{4\to 2}^w -k_{4\to 3}^w
\end{bmatrix}
\end{equation}
\end{widetext}
The rate constants are 
\bea
k_{1\to 2}^{c}&=&\frac{1}{2}J_{c}(\theta_c^-) n_{c}(\theta_c^-),\quad k_{1\to 3}^{c}=\frac{1}{2}J_{c}(\theta_c^+) n_{c}(\theta_c^+)
\nonumber\\
k_{2\to 4}^{w}&=&\frac{1}{2}J_{w}(\theta_w^+) n_{w}(\theta_w^+),\ \ k_{3\to 4}^{w}=\frac{1}{2}J_{w}(\theta_w^-) n_{w}(\theta_w^-)
\nonumber\\
k_{1\to 4}^{h}&=&J_{h}(\theta_h) n_h(\theta_h),
\eea
with the  reversed rate constants determined from detailed balance.
As in Appendix A, the factor $1/2$ results from the $1/\sqrt2$ prefactor in front of the operator $\hat S$, Eq. (\ref{eq:S4l}).
We use an ohmic spectral density with a large cutoff. As well, for simplicity, we assume a symmetric coupling,
$\gamma=\gamma_{h,c,w}$. This leads to, e.g.,
$k_{1\to 2}^{c}=\frac{1}{2}\gamma \theta_c^- n_{c}^-$,  with
$n_{\alpha}^{\pm}\equiv n_{\alpha}(\theta_{\pm})$. 
We calculate the energy currents using Eq. (\ref{eq:Jq}) and get
\bea
J_{q}^c &=& \frac{\theta_c^+}{\Phi}(k_{2\to 1}^c+k_{2\to 4}^w)
(k_{1\to 3}^ck_{3\to 4}^wk_{4\to 1}^h-
k_{3\to 1}^ck_{4\to 3}^wk_{1\to 4}^h)
\nonumber\\
\nonumber\\
&+&\frac{\theta_c^-}{\Phi}(k_{3\to 1}^c+k_{3\to 4}^w)
(k_{1\to 2}^ck_{2\to 4}^wk_{4\to 1}^h-
k_{2\to 1}^ck_{4\to 2}^wk_{1\to 4}^h)
\nonumber\\
&-&\frac{2g}{\Phi}
(k_{1\to 2}^ck_{3\to 1}^ck_{2\to 4}^wk_{4\to 3}^w
-k_{1\to 3}^ck_{2\to 1}^ck_{3\to 4}^wk_{4\to 2}^w).
\label{eq:CJq}
\eea
%
This expression agrees with Ref. \cite{jose15}.
Here, $\Phi$ is a 
positive normalization factor which is quite cumbersome and has been left out.
The first two lines in Eq. (\ref{eq:CJq}) describe heating and cooling processes through the two transitions $\theta_{\pm}$.
The last line corresponds to heat leaks, with thermal energy exchanged directly between the work and cold baths.
Similarly to the V-system, even in the secular limit the behavior of the energy current is highly non-monotonic in $g$, 
affecting the magnitude of cooling current and the cooling window.

Taking $g\to 0$, the mid-gap states in the 4lQAR model become degenerate and we realize a setup that 
looks and behaves very much like the 3lQAR. In this limit, the current reduces to
\bea
J_{q}^c=2\theta_c \frac{k_{1\to 2}^ck_{2\to 4}^wk_{4\to 1}^h-k_{1\to 4}^hk_{2\to 1}^ck_{4\to 2}^w}{\Psi},
\label{eq:CJqsmall}
\eea
with 
\bea
\Psi&=&
k_{1\to 4}^hk_{2\to 1}^c+2k_{1\to 2}^ck_{2\to 4}^w+2k_{1\to 2}^ck_{4\to 1}^h+k_{1\to 4}^hk_{2\to 4}^w 
\nonumber\\
&+& 
k_{2\to 1}^ck_{4\to 1}^h
+4k_{1\to 2}^ck_{4\to 2}^w+2k_{1\to 4}^hk_{4\to 2}^w 
\nonumber\\
&+& 2k_{2\to 1}^ck_{4\to 2}^w+k_{2\to 4}^wk_{4\to 1}^h.
\eea
This result is analogous to the current for the 3lQAR, Eq. (\ref{eq:Jq_3lQAR})---with a different denominator 
(normalization factor), and a factor of 2 here, accounting for the two effective cycles.
Note that level `3' in the 3lQAR model serves as level `4' in the 4lQAR model.
Also note that the rate constants defined for the 4lQAR are half the value of the rate constants defined for the 3lQAR.

In this $g\to 0$ and large $\gamma_d$ limit, we find that  the cooling current of the 4lQAR, Eq. (\ref{eq:CJqsmall}), 
is smaller than the one for the 3lQAR, Eq. (\ref{eq:Jq_3lQAR}),
but the cooling window is identical, Eq. (\ref{eq:cond}).
One can similarly calculate the thermal energy absorbed from the work bath, $J_q^w$, and find that
it follows Eq. (\ref{eq:CJqsmall}), only with the energy prefactor $\theta_w$.
Thus, when $g\to 0$  and $\gamma_d$ is large, the coefficient of performance for the 4lQAR is identical to that
of the  3lQAR , $\eta=\theta_c/\theta_w$.


\end{document}